\documentclass[onecolumn,english,aps,showpacs,superscriptaddress,notitlepage]{revtex4-1}
\usepackage{amsmath,amssymb}
\usepackage{graphicx}
\usepackage{txfonts}
\usepackage{color}
\usepackage{hyperref}
\usepackage{bbold}
\usepackage{framed}
\usepackage{wrapfig}
\usepackage{multirow}
\usepackage[letterpaper]{geometry}
\global\long\def\av#1{\left\langle #1 \right\rangle }

\global\long\def\Tr{{\rm Tr}}

\renewcommand{\vec}[1]{\boldsymbol{#1}}
\newcommand{\beq}{\begin{equation}}
\newcommand{\eeq}{\end{equation}}
\newcommand{\bea}{\begin{eqnarray}}
\newcommand{\eea}{\end{eqnarray}}

\begin{document}

\title{Nematic superconductivity in twisted bilayer graphene}

\author{Dmitry V. Chichinadze}
\author{Laura Classen}
\author{Andrey V. Chubukov}
\affiliation{School of Physics and Astronomy,
University of Minnesota, Minneapolis, MN 55455, USA}
\begin{abstract}
Twisted bilayer graphene displays insulating and superconducting phases caused by exceptional flattening of its lowest energy bands.
Superconductivity with highest $T_c$ appears at hole and electron dopings,
near  half-filling for valence or conduction bands.
In the hole-doped case, the data show that three-fold lattice rotation symmetry is broken in the superconducting phase, i.e., a superconductor  is also a nematic.
We perform a comprehensive analysis of superconductivity in twisted-bilayer graphene within an itinerant approach and  present a mechanism for nematic superconductivity.
We take as an input the fact that at dopings, where superconductivity has been observed,
 the Fermi energy lies in the vicinity of twist-induced Van Hove singularities in the density of states.
We argue that the low-energy physics can be properly described by patch models with six Van Hove points for electron doping and twelve Van Hove points for hole doping.
We obtain pairing interactions for the patch models in terms of parameters of the microscopic model for the flat bands, which contains both local and  twist-induced non-local interactions.
   We show that the latter give rise to attraction in different superconducting channels.
For electron doping, there is just one attractive $d-$wave channel,  and we find chiral $d\pm id$ superconducting order, which breaks time-reversal symmetry, but leaves the lattice rotation symmetry intact.
For hole-doping, we find two attractive channels, $g$ and $i$-waves, with almost equal coupling constants.
We show that both order parameters are non-zero in the ground state, and explicitly demonstrate
  that in this co-existence state
 the threefold lattice rotation symmetry is broken, i.e., a superconductor is also a nematic.
 We find two possible nematic states, one is time-reversal symmetric, the other additionally breaks time-reversal symmetry.
Our scenario for nematic superconductivity is based on generic symmetry considerations, and we
 expect it to be applicable also to other systems with two (or more) attractive channels with similar couplings.
\end{abstract}

\maketitle

\section{Introduction}

Twisted hexagonal heterostructures recently joined the family of condensed matter systems which display superconductivity with yet unsettled pairing mechanism.
Superconductivity (SC) has been observed in twisted bilayer graphene \cite{Cao2018unconventional,Cao2018correlated,Yankowitz2019,Lu2019}, twisted double-bilayer graphene \cite{2019arXiv190308596C,2019arXiv190308130L,2019arXiv190306952S}, and trilayer graphene on boron nitride \cite{trilayer1,trilayer2} under various tuning conditions controlled by twist angle, pressure, filling, or external field.
This  high degree of tunability holds the promise to
enlighten the pairing problem from many different angles and thereby improve our understanding
of superconductivity in correlated electron systems.
 Like in several other correlated systems, SC borders insulated phases, in which fermions either get localized by Mott physics, or develop a competing order, which gaps excitations near the Fermi surface.

The occurrence of SC and insulating phases is ascribed to an exceptional band flattening, which comes along with a very large hexagonal moir\'e pattern in real space \cite{Suarez2010,Bistritzer2011}.
The small bandwidth of the resulting isolated flat band increases the relative strength of both electron-electron \cite{Po2018PRX,Kang2018strong,Koshino2018PRX,Seo2019}  and  electron-phonon \cite{WuSDS,WuMCD}  interaction and promotes correlation effects \cite{Wolf2019}.
In twisted bilayer graphene  (TBG), insulating and superconducting phases have been induced by adjusting the band flatness either by changing the twist angle between the two graphene layers
to the so-called magic angle~\cite{Bistritzer2011} close to 1.1$^\circ$, or upon applying pressure in the vicinity of the magic angle  \cite{Cao2018unconventional,Cao2018correlated,Yankowitz2019,Lu2019}.
 The most prominent insulating states occur near half filling of either conduction or valence bands at the density of two electrons or two holes  per moir\'e unit cell  (near  $n=\pm 2$ in the classification where $n=4$ corresponds to fully occupied and $n =-4$ to empty flat bands).
 Insulating  states have also been reported near other integer fillings
     \cite{Yankowitz2019,Lu2019}.   Similarly, the SC state with the highest critical temperature $T_c \sim 3K$ \cite{Lu2019} was detected close to half-filling of the valence band ($n =-2$).
         Further superconducting states with  $T_c \leq 0.65K$ have been found near the half-filled conduction band and in between other integer fillings.
The suppression of superconductivity by a  small magnetic field points to spin-singlet pairing \cite{Cao2018unconventional}.

The data indicate~\cite{stm2019} that TBG with a twist angle near the magic one lies in a regime of moderate coupling, where the ratio of Coulomb interaction (reduced by the large spatial scale of the moir\'e pattern) and the width of the flat band is of order one (both are in the range of $10-20$ meV).
  Consequently,  arguments have been made for both  moderate coupling itinerant approach
   and  strong coupling Mott-type approach.  Arguments for the strong coupling approach have been rationalized by the fact that insulating states have been detected  not only near $n = \pm 2$, but also near other integer fillings $n = \pm 1$ and $\pm 3$.
Arguments for an itinerant approach are based on the observations that even the highest superconducting $T_c$ of 3K is much smaller than the bandwidth, and that insulating behavior is rather fragile -- it disappears already for small $T \geq 10K$ and for small fields $H \leq 5T$, Ref. \cite{Cao2018correlated,Yankowitz2019,Lu2019}.
 Within  the itinerant approach, insulating states are viewed as competing states with some type of order in the particle-hole channel, and superconductivity and competing orders are largely based on the notion that close to half-filling \cite{vhs2016,stm2019} and, possibly,  other integer fillings,  the chemical potential nearly coincides with twist-induced Van Hove (VH) singularities  in the density of states, Ref.~\cite{e_andrei}.
This generally amplifies the effect of interactions both in particle-hole and particle-particle channels \cite{HUR20091452}.

  In our study, we analyze superconductivity within an  itinerant approach.
 Earlier studies considered both phonon~\cite{Samajdar2020microscopic,PhysRevLett.122.257002,PhysRevB.98.241412,PhysRevB.98.220504} and purely electronic pairing mechanisms \cite{Venderbos2018,Isobe2018PRX,Ray2019,Gonzalez2019,PhysRevB.98.205151,You2019npj,Lin2018,Lin2019,2019arXiv190301701C,2018arXiv180506906W,PhysRevB.98.075154,PhysRevLett.121.087001,
 PhysRevB.98.085436,LAKSONO201838,PhysRevLett.121.217001,PhysRevB.98.195101,Liu_2019,PhysRevB.98.241407,PhysRevB.99.195120}.  The works on the electronic mechanism often explored a scenario,   where the enhancement of the pairing in the doubly degenerate $d-$wave channel at densities near the  VH singularities
  leads to chiral $d \pm id$ superconductivity.  A similar  analysis had been previously  performed for a single layer of graphene at high doping \cite{Nandkishore2012,PhysRevB.86.020507}.

 The primary goal of our study is to take a new look at superconductivity in the presence of VH singularities near the Fermi level in view of the recent experimental observations that a discrete $C_3$  rotational symmetry is likely broken in the superconducting state of hole-doped TBG ~\cite{kitp_talk}.
  Strong nematic fluctuations
  were  earlier reported in the STM measurements in the normal state~\cite{stm2019,Jiang2019}.
  The authors of~~\cite{kitp_talk} argued that for some hole dopings  $C_3$ breaking  extends to the normal state, but for other dopings, where superconductivity has been observed, $C_3$ breaking is only present below $T_c$.
   The direction of the nematic order is different at dopings where nematicity is only seen in the superconducting state and where it extends to the normal state. This likely indicates that the nematicity in the normal state  is not the source of the nematic order at dopings where it emerges below $T_c$.  We take these results as a motivation for our study and analyze a possibility to obtain $C_3$ breaking only in the superconducting state.

\begin{figure}[tb]
\includegraphics[width=0.5\linewidth]{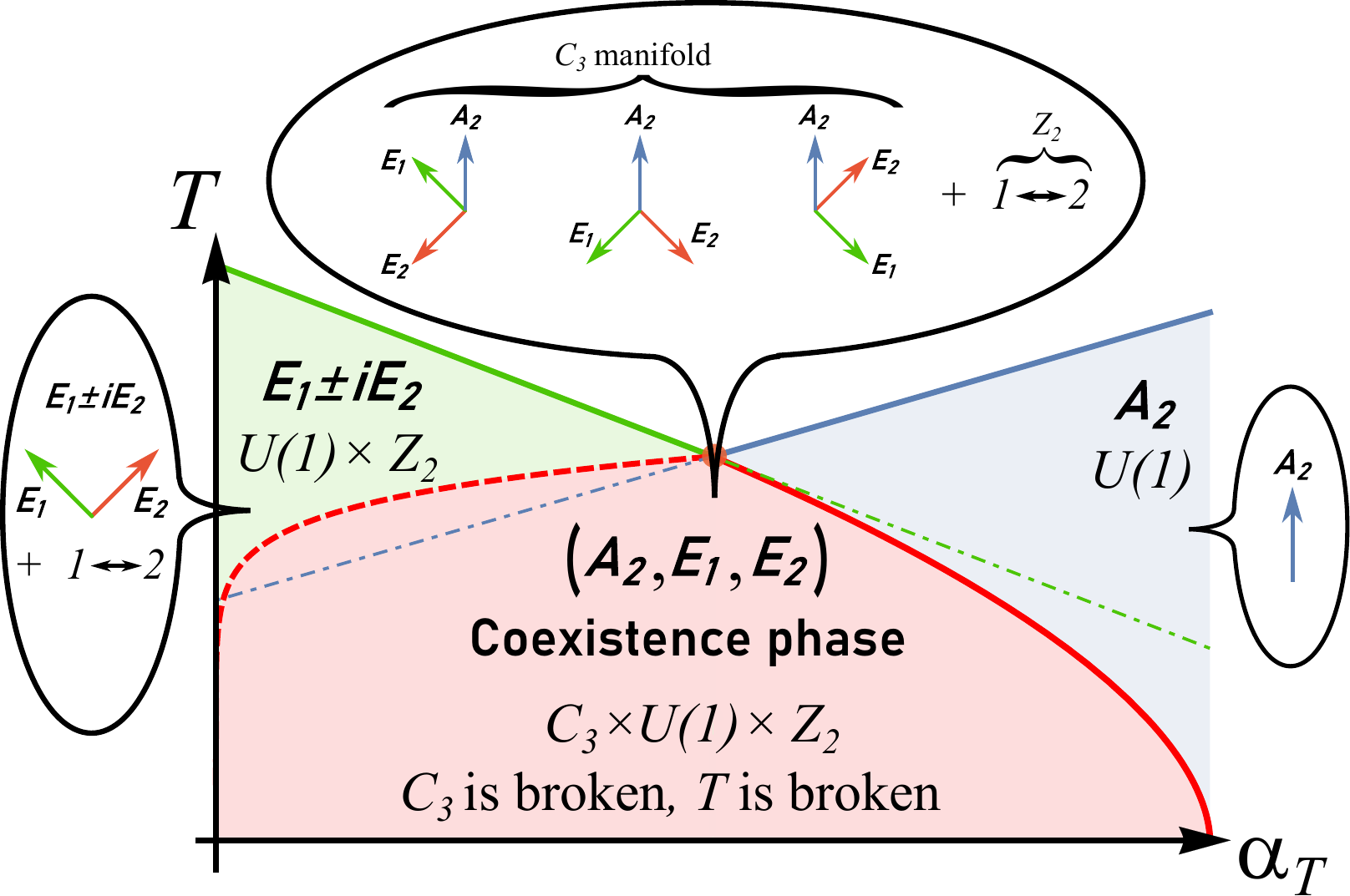} \quad
\includegraphics[width=0.43\linewidth]{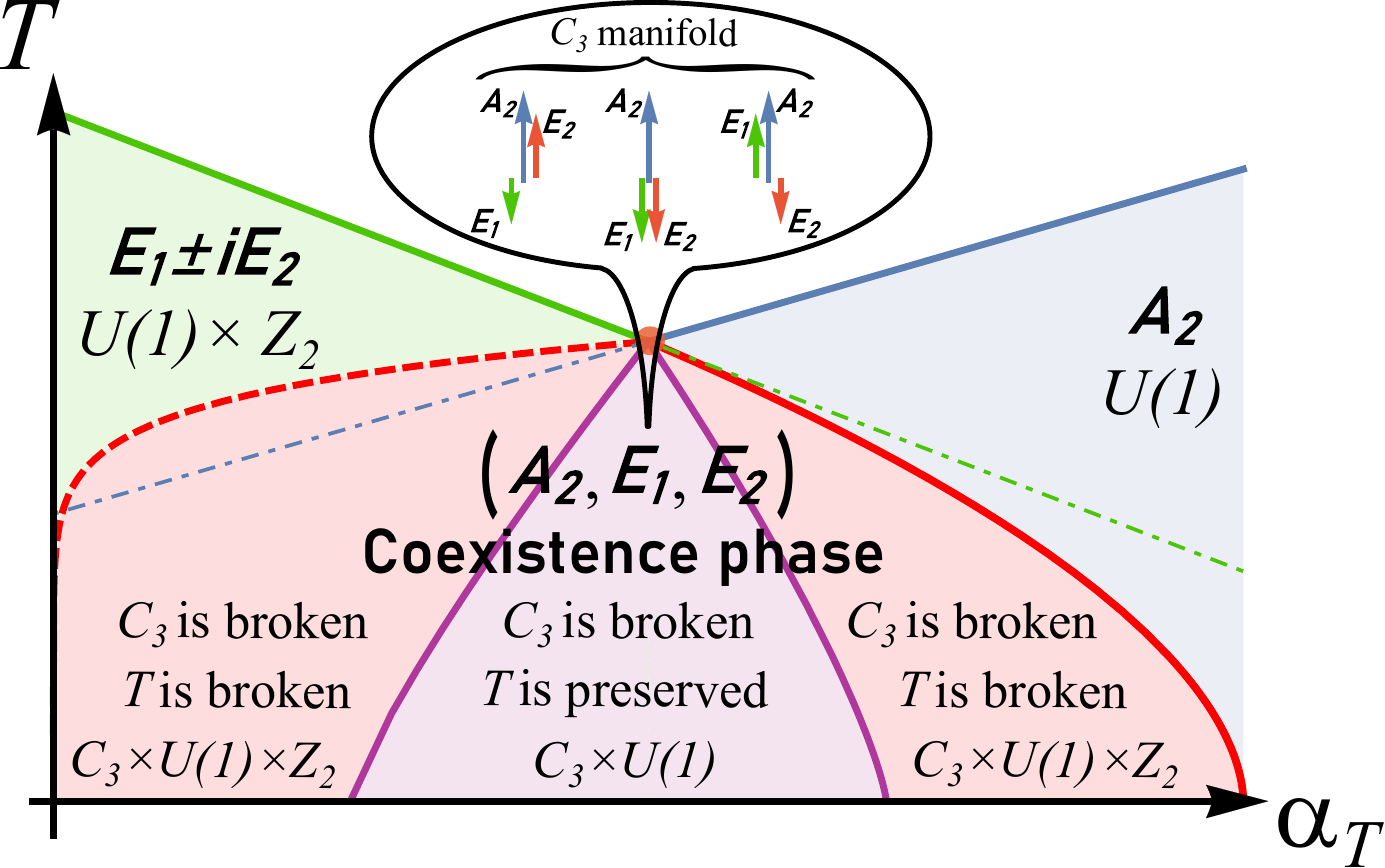}
\centering{}\caption{
A schematic phase diagram for the case when the pairing channels with $E$ and $A_2$ symmetry are nearly degenerate (i.e., the attractive interaction in the two channels has almost the same magnitude).
 We use the strength of the non-local interactions $\alpha_T$ as a parameter to distinguish between the cases when
 the interaction in one of the channels is  stronger than in the other.
  Immediately below the onset of the pairing, SC develops in one of the two channels, and the gap symmetry is
    either $g\pm i g$  ($E$-channel, green shaded region) or $i$  ($A_2$-channel, blue shaded region), both states are rotationally symmetric.
At a lower temperature,  a coexistence state develops (red shaded region).
Depending on the parameters in the Landau free energy, this state is either type I or type II.
State of type I (left panel) evolves continuously between pure $A_2$ and pure $E$ states.
 This state breaks  $C_3$ rotational symmetry, i.e., it is a nematic superconductor, and also breaks time-reversal symmetry.
  For state of type  II (right panel), there is an additional intermediate phase inside the coexistence region (purple region).  In this intermediate phase, $C_3$ rotational symmetry is broken, but time-reversal symmetry is preserved.
 In both panels, the transition between  the $E$ state and the co-existence state is first order (red dashed line), and the transition between the co-existence state and the $A_{2}$ state is second order (solid red line).
The three states in the "bubble" in the middle of each panel are symmetry partners of the $C_3$ manifold;
 time-reversal symmetry is realized by exchanging $E_1\leftrightarrow E_2$.
The directions of blue, green, and brown arrows correspond to the phases of $A_2, E_1,$ and $E_2$ order parameters, $\Delta_{i} = |\Delta_{i}| e^{i \phi_{i}}$, counted from the $y$ axis.  For definiteness, we set the $A_2$ order parameter to be real.
}
\label{phdiag1}
\end{figure}

   The breaking of a lattice rotational symmetry is usually associated with nematic order,  and a state in which this symmetry is broken below the superconducting $T_c$ is called  nematic superconductor.
Nematic superconductivity has been earlier discussed for LiFeAs (Ref. \cite{borisenko}) and doped topological insulator Bi$_2$Se$_3$ (Refs. \cite{PhysRevLett.104.057001,PhysRevLett.105.097001,Fu2014odd,Matano2016,Venderbos2016odd,Yonezawa2017,Schmalian2018,2019arXiv190501702C}).
We argue that the scenarios proposed for these materials do not apply to TBG.
The electronic scenario, discussed in earlier works,
 yields $d \pm id$ superconducting order
~\cite{You2019npj,Lin2018,Lin2019,2019arXiv190301701C,PhysRevLett.121.087001,PhysRevB.98.085436,PhysRevLett.121.217001,PhysRevB.98.241407,PhysRevB.99.195120}. This order breaks time-reversal symmetry, but preserves $C_3$ rotational symmetry. 
Phonon-mediated pairing with $s-$wave superconductivity does not break $C_3$ symmetry either.
Here, we propose a novel scenario for pairing in TBG, which gives rise to  nematic superconductivity.
 We argue that a combination of the geometry of VH points near $n =-2$, which double in number compared to
electron doping, and the form of the effective interaction, which was argued to possess both on-site and nearest-neighbor components \cite{Koshino2018PRX,Kang2018strong,Seo2019}, gives rise to an attraction in two pairing channels.
We note that, as soon as the two pairing channels are attractive, our analysis is based on symmetry and independent on the pairing mechanism leading to the attraction.
 In the classification of the lattice rotation group $D_3$, appropriate for TBG,  one of the attractive channels is the doubly degenerate $E$ channel and the other is the single-component $A_2$ channel.
The $d$-wave state,  discussed  in earlier works,  belongs to $E$ channel.
Based on the number of nodes along the Fermi surface,  the SC states that we found here correspond to doubly degenerate "$g$-wave" and "$i-$wave",  respectively (each gap component in the $E$ channel changes sign eight times under a $2\pi$ rotation around the center of the Brillouin zone, while the gap in the $A_2$ channel changes sign twelve times under a full rotation).
The presence of higher lattice harmonics in the superconducting gap structure of TBG was also discussed in \cite{Wu2019harmonic}.
We argue that the coupling constants in the two channels are quite close, so in a wide temperature range below $T_c$ the system is in the coexistence state, where both SC orders are present.  Taken alone, each of the two states  does \emph{not} break $C_3$ symmetry: the order parameter in the $E$ channel $E_1 \pm i E_2$ (the $g$-wave analogue of $d\pm id$) breaks $U(1)$ phase and $Z_2$ time-reversal symmetry, and  the order parameter in the $A_2$ channel breaks $U(1)$ phase symmetry.  We show that in the coexistence state, $C_3$ symmetry is broken, along with the overall $U(1)$  phase symmetry.
The breaking of $C_3$ symmetry is due to the presence of special coupling terms in the Landau
free energy, which are linear in the $A_{2}$ order parameter and cubic in the $E$ order parameter.
 We found two coexistence states: in one time-reversal symmetry is additionally broken, in the other it is preserved.
 The time-reversal-symmetric state develops if the special coupling between the two order parameters is sufficiently strong.
 Otherwise, time-reversal is broken in the coexistence state.
 We show the corresponding phase diagrams in Fig. \ref{phdiag1}
 as function of a tuning parameter $\alpha_T$, which determines the non-local interaction strength and  regulates how close the two different SC states are in energy.

It is instructive to compare our findings with  the
 earlier proposals
 for nematic superconductivity in TBG.  For spin-singlet pairing, earlier works~\cite{Venderbos2018,Kozii2019} focused on the two-component $E$ state, without an admixture of the $A_2$ state.
 In this situation, nematic superconductivity can develop if the solution for the gap is non-chiral, $(\Delta_{E_1}, \Delta_{E_2}) = \Delta_E (\cos{\gamma}, \sin{\gamma})$, and the minima of the free energy  are at three values of $\gamma$.
 We looked into this possibility, but found that for parameters extracted from the microscopic model \cite{Yuan2018,Koshino2018PRX,Kang2018strong} that we use, the solution for the $E$ state is the chiral  $\Delta_{E_1} \pm i \Delta_{E_2}$, which breaks time-reversal, but preserves $C_3$ lattice rotational symmetry.
 It was suggested~\cite{Kozii2019}  that fluctuation corrections  can  potentially change the free energy  and make the nematic configuration energetically favorable, if the
 nematic component of density wave fluctuations is large in the normal state.
 In a similar spirit, it was argued in  Ref.~\cite{PhysRevB.98.075154} that fluctuation-induced  nematic
  superconductivity can develop in the vicinity of a transition into a nematic orbital ferromagnet.
   We did not analyze fluctuation corrections or  preformed nematic phases in our model.
   Instead, we focus on the superconductivity coming already from the bare interaction.
   In Ref.~\cite{Scheurer2019} nematic superconductivity in the triplet channel has been explored.
  This work is  likely applicable to twisted double-bilayer graphene, where data suggest
  spin-polarized pairing \cite{,2019arXiv190308130L}.
   In TBG, which we consider,  experiments point to spin-singlet pairing \cite{Cao2018unconventional}.

    The doubling of the number of Van Hove points as function of twist angle or pressure, and the difference in the number of VH points  in valence and conduction bands has been  considered for  twisted bilayer graphene in Refs. \cite{Gonzalez2019,Yuan2019magic} and for  monolayer jacutingaite in Ref. \cite{PhysRevB.100.041117}.
    In particular, Ref.~\cite{Gonzalez2019} analyzed Kohn-Luttinger superconductivity within the model with twelve VH points and Hubbard interactions.  They found attraction in
    several channels, with the dominant one being spin-triplet and $C_3$ symmetric.
     Our analysis differs from Ref.~\cite{Gonzalez2019} in two aspects. First, we argue that the non-local component of the interaction gives rise to an attraction in more that one channel already at the bare level, and, second, we argue that, to find nematic superconductivity, one needs to move below the highest $T_c$ and analyze the coexistence phase.
We also argue that the difference in the number of VH points in twisted bilayer graphene (six for electron doping  vs twelve for hole doping)  leads to different SC states, and that nematic superconductivity develops only for hole doping.

The structure of our paper is as follows. In Sec.~\ref{sec:TBM}  we introduce the patch models with six and twelve VH points and extract the model parameters from the underlying microscopic tight-binding model with local and non-local interactions.
 In Sec.~\ref{sec:gap} we solve the corresponding linearized gap equation and determine the symmetries of its solutions.
We use the result to derive the Landau free energy for spin-singlet superconductivity in TBG in Sec.~\ref{sec:GL}. We analyze the free energy in detail and present possible SC phase diagrams in the end. We conclude in Sec.~\ref{sec:concl}.
  Several details of the derivations are presented in the Supplementary material.

Before we proceed, we present a brief summary of our results.

\subsection*{Summary of the results}

 We use as an input for our study
   the effective tight-binding model for the flat bands introduced in Refs.~\cite{Yuan2018,Koshino2018PRX,Kang2018PRX,Kang2018strong}.
   We consider fermionic densities at which VH singularities are located near the chemical potential, and
    introduce patch models for fermions in hot regions near the VH points.  We argue that the proper patch model near  $n \approx 2$  contains six patches, and the proper patch model near $n =-2$ contains twelve patches.
  We consider all symmetry-allowed interactions between hot fermions and extract their values from
   the microscopic model of Ref.  ~\cite{Kang2018strong}. In this model the interaction term has both  local (Hubbard) and non-local (nearest-neighbor) components, with comparable
    strength.

Within the patch models, we solve for spin-singlet pairing in various channels. The pairing states can be classified according to irreducible representations of the point group $D_3$, appropriate for TBG.
This point group has three irreducible representations: two one-dimensional representations $A_1, A_2$ and one two-dimensional representation $E$ \cite{Hamermesh}.
  We find that the interaction in some pairing channels is attractive, because of the non-local component.

For the six-patch model near $n=2$ we find that the $E$-channel ($d$-wave) is attractive, while the $A_1$ channel ($s$-wave) is repulsive. The $A_2$ channel ($f$-wave) does not contribute to spin-singlet pairing.
This is in accordance with earlier studies of the six-patch model for TBG~\cite{Isobe2018PRX} and single-layer graphene~\cite{Nandkishore2012}.
The corresponding SC  order parameter can be represented by a vector ${\bf \Delta}^{E}_{6p}$ with number of components equal to
the number of patches.
 Because the $E$ channel is two-dimensional, there are two independent order-parameter vectors ${\bf \Delta}^{E_1}_{6p}$  and ${\bf \Delta}^{E_2}_{6p}$, and the full SC gap is a linear combination $\Delta^{SC}_{6p}  = \Delta_{E_1} {\bf \Delta}^{E_1}_{6p} + \Delta_{E_2} {\bf \Delta}^{E_2}_{6p}$.
The type of the superconducting order depends on which linear combination minimizes the free energy. It is determined by the sign of the coupling term between $\Delta_{E_1}$  and $\Delta_{E_2}$:  $\beta_2 |\Delta^2_{E_1}  + \Delta^2_{E_2}|^2$. When $\beta_2 >0$, a chiral $\Delta_{E_1} \pm i \Delta_{E_2}$ state develops, when $\beta_2 <0$ nematic SC  develops with $(\Delta_{E_1}, \Delta_{E_2}) = \Delta_E (\cos{\gamma}, \sin{\gamma})$ .
 For the  microscopic model that we use for TBG, we find $\beta_2>0$, i.e. the SC state is chiral $\Delta_{E_1} \pm i \Delta_{E_2}$.
 This order breaks time-reversal symmetry, but preserves lattice rotational symmetry (the gap amplitude is the same at all six VH points).

   For the twelve patch model near $n =-2$, we find that two channels, $E$ and $A_2$, are attractive with nearly equal coupling constants.
   Analogously to the six-patch case, the corresponding order parameters are twelve-component vectors,
   which can be expressed as $\Delta^{SC}_{12, E}= \Delta_{E_1} {\bf \Delta}^{E_1}_{12p}  + \Delta_{E_2} {\bf \Delta}^{E_2}_{12p}$ and $\Delta^{SC}_{12, A_2}= \Delta_{A2} {\bf \Delta}^{A_2}_{12p}$.
 The minimization of the free energy for the $E$ state taken alone again yields the chiral $\Delta_{E_1} \pm i \Delta_{E_2}$ state that breaks time-reversal symmetry, but preserves $C_3$ lattice rotation symmetry.
 The  $A_2$ state with a single gap amplitude $\Delta_{A_2}$  preserves both time-reversal and lattice rotation symmetries.
  However, we show that new states emerge at low temperatures, when both  $E$ and $A_2$ gaps are non-zero.  To study the order parameter in the coexistence state, we derive the Landau functional
   $F[\Delta_{E_1}, \Delta_{E_2}, \Delta_{A2}]$.  The functional, taken to quartic order in $\Delta_i$, contains regular mixed terms, quadratic in $\Delta_{E_1}, \Delta_{E_2}$, and in $\Delta_{A_2}$: $\gamma_1 (|\Delta_{E_1}|^2 + |\Delta_{E_2}|^2) |\Delta_{A_2}|^2 + \gamma_2 \left((\Delta^2_{E_1} + \Delta^2_{E_2}) {\bar \Delta}^2_{A_2} + c.c\right)$,  and the asymmetric term
   $\delta \bar\Delta_{A_2}\left[ (\Delta_{E_1}-i \Delta_{E_2})^2(\bar\Delta_{E_1}-i \bar\Delta_{E_2})+ (\Delta_{E_1}+i \Delta_{E_2})^2(\bar\Delta_{E_1}+i \bar\Delta_{E_2}) \right]+$ c.c.
  The  coefficients $\gamma_{1,2}$, $\beta_2$, and $\delta$ are all expressed via the parameters of the underlying microscopic model.
  The asymmetric term is special in the sense that it is linear in $\Delta_{A_2}$ and qubic in $\Delta_{E_{1,2}}$, yet it is invariant under all symmetry transformations from the $D_3$ space group on the hexagonal lattice,
as well as under time-reversal and $U(1)$ gauge transformations.
We argue that because of  this term, the order parameter in the coexistence state breaks $C_3$ lattice rotational symmetry.

To illustrate the root of the $C_3$ breaking, we analyze separately the special case when the asymmetric term is absent, and the generic, proper case when it is present. In the  special  case, we found that there are two coexistence states. Both are highly degenerate, with order parameter manifold $U(1) \times U(1) \times Z_2$ in one phase, and $U(1) \times  U(1)$ in the other.  The presence of two $U(1)$'s implies that there is an additional continuous degeneracy besides the conventional $U(1)$ overall phase  degeneracy.
   The extra $Z_2$ in one phase is associated with  time-reversal. In the other phase, time-reversal operation is a part of the extra $U(1)$ symmetry.
 In a generic case, when $\delta$ is finite, we find that the additional $U(1)$ gets discretized.
  For small $\delta$, we find that there exists a single coexistence phase with order parameter manifold $U(1) \times C_3 \times Z_2$, where $U(1)$ is phase degeneracy, $C_3$ is a discrete symmetry with respect to lattice rotations,  and  $Z_2$ is associated with time-reversal.  The superconducting order breaks all three symmetries, including $C_3$ symmetry of lattice rotations.
      This implies that the  coexistence state is a nematic superconductor.
    For larger $\delta$ we find that there appears a region within the coexistence state, where the order parameter manifold is $U(1) \times C_3$.  A SC order in this range is again  nematic, but it  does not break time-reversal symmetry.
 For the parameters of the microscopic model of Refs.~\cite{Yuan2018,Koshino2018PRX,Kang2018strong}, the value of $\delta$ is close to the boundary where the state with broken $C_3$ and unbroken time-reversal symmetry develops.  We therefore cannot rigorously argue for or against time-reversal breaking in the superconducting state of TBG.  Still,  we emphasize that for any $\delta >0$, the SC state  in our twelve-patch model  near $n =-2$  breaks $C_3$ lattice rotational symmetry, i.e., the SC state is also a nematic state.  This is consistent with the experiments, which near $n =-2$  found a two-fold anisotropy of resistivity in the vortex state as function of the direction of the applied magnetic field~\cite{kitp_talk}.

\section{Effective patch models from tight-binding}
\label{sec:TBM}

\subsection{Fermiology of twisted bilayer graphene}

While a brute-force microscopic description of TBG is obstructed by the huge unit cells of the moir\'e superlattice, low-energy continuum \cite{PhysRevLett.99.256802,PhysRevB.81.161405,PhysRevB.84.235439,Bistritzer2011,PhysRevB.86.155449,Moon2012,Moon2013,vhs2016,Koshino2015JPSJ,Nam2017} and tight-binding \cite{Shallcross2010,Suarez2010,Trambly2010,PhysRevB.89.205414,Fang2016} models that couple both layers have been very successful in analyzing the electronic properties of TBG -- including the theoretical prediction of flat bands itself ~\cite{PhysRevLett.99.256802,Trambly2010,PhysRevB.86.155449,Bistritzer2011}.
  More recent works derived effective tight-binding models for the superlattice based on localized Wannier states exclusively for the isolated flat bands \cite{Yuan2018,Koshino2018PRX,Kang2018PRX,Po2018PRX}.
 These Wannier states have a three-peak structure centered around sites of the honeycomb lattices, which is dual to the triangular moir\'e lattice where the local charge density is concentrated. Such a structure gives rise to hopping between further neighbors.
In this context, the ability to write down a tight-binding model exclusively for TBG flat bands has been discussed. To do so, one has to overcome Wannier obstructions \cite{PhysRevB.99.195455,Zou2018PRB}. The obstruction occurs if one implements symmetries at incommensurate twist angles which are not exact, but assumed to emerge. To avoid the obstruction, but still construct a tight-binding model for the flat-bands only, one can either consider commensurate structures near the magic angle with well-defined bands \cite{Kang2018PRX}, or sacrifice one of the approximate symmetries \cite{Koshino2018PRX}. This is why the model we use below has a three-fold symmetry instead of a six-fold one.

\begin{figure}[t]
\center{\includegraphics[width=0.2\linewidth]{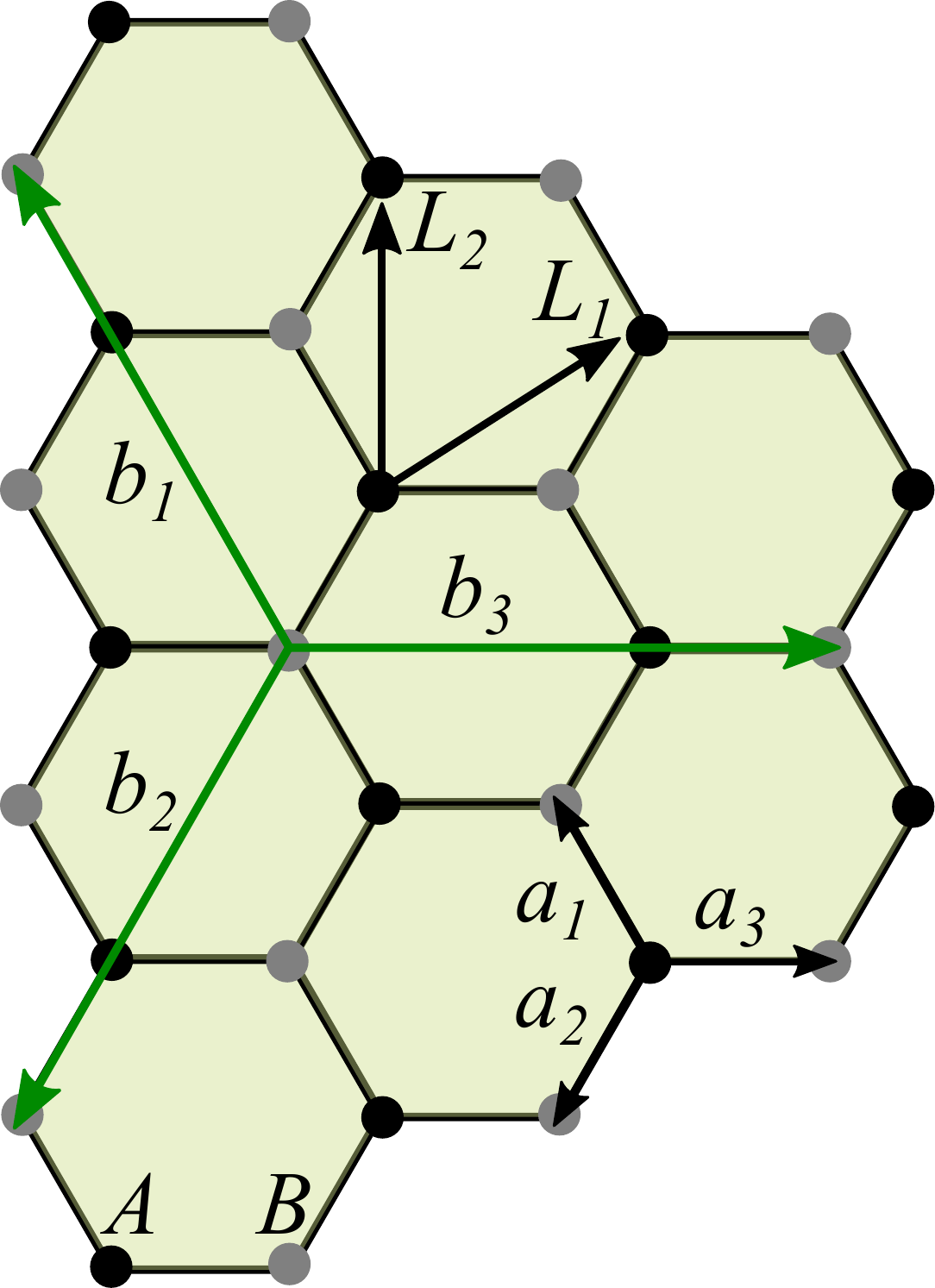} }
\centering{}\caption{The moir\'e honeycomb superlattice and translation vectors relevant for the tight-binding Hamiltonian. Vectors $\vec{a}_i$ correspond to nearest-neighbor hopping, $\vec{b}_i$ correspond to  hopping between fifth nearest neighbors (denoted by $\av{ij}_5$ in Eq. (\ref{2H})), and $\vec{L}_i$ are primitive lattice vectors. Black and gray circles show $A$ and $B$ sublattice sites, respectively.}
\label{lattice}
\end{figure}

In this work, we employ the model for the dispersion proposed by Yuan and Fu \cite{Yuan2018} (see also \cite{Koshino2018PRX}).
We start from writing down the tight-binding Hamiltonian for the moir\'e superlattice in real space in terms of Wannier states
\begin{align}\label{2H}
H_{TB} &= H_{TB}^{(0)} + H_{TB}^{(1)} \\
H_{TB}^{(0)} &= - \sum_i \mu \bold{c}_i^{\dagger} \bold{c}_i + \sum_{\av{ij}} t_1 \left[\bold{c}_i^{\dagger} \bold{c}_j + h.c. \right] +  \sum_{\av{ij}_5} t_2 \left[\bold{c}_i^{\dagger} \bold{c}_j + h.c. \right], \\
H_{TB}^{(1)} &=  \sum_{\av{ij}_5} t_3 \left[ \left(\bold{c}_i^{\dagger} \times \bold{c}_j \right)_z + h.c. \right].
\end{align}
Here, the sums go over the sites of the honeycomb lattice, which are centered on the AB or BA regions of the moir\'e pattern in TBG.  The operators $\bold{c}_i = \left(c_{i,x}, c_{i,y} \right)^T$ annihilate electrons with  $p$-wave-like orbital index $x$ and $y$, $\mu$ denotes the chemical potential, $t_1, t_2$ are real hopping amplitudes between nearest- and fifth-nearest-neighbors, and $\av{ij}_5$ denotes fifth-nearest neighbor (see Fig. \ref{lattice}). A fifth-nearest neighbor is equivalent to a second-nearest neighbor within the same sublattice.   For simplicity, we suppressed a spin index.

The Hamiltonian  $H_{TB}^{(1)}$ possesses an orbital and spin $U(1) \times SU(2)$ symmetry, $D_3$ space symmetry of the TBG lattice, and is symmetric under time reversal.
It yields four spin-degenerate bands with dispersions
\begin{equation}
E^\pm_\pm = T_d \pm T_{sd2} \pm \sqrt{|T_{sd1}|^2} ,
\label{band_spec}
\end{equation}
where
\begin{align}
T_d &=-\mu+2 t_2 (\cos{\vec{b}_1 \vec{k}} + \cos{\vec{b}_2 \vec{k}} +\cos{\vec{b}_3 \vec{k}}), \\
T_{sd1} &= t_1 \left( \exp(i k_x) + 2 \exp(-i \frac{k_x}{2}) \cos({\frac{\sqrt{3} k_y}{2}}) \right), \\
T_{sd2} &= 2 t_3 (\sin{\vec{b}_1 \vec{k}} + \sin{\vec{b}_2 \vec{k}} +\sin{\vec{b}_3 \vec{k}}).
\end{align}

\begin{figure}[t]
\center{\includegraphics[width=0.4\linewidth]{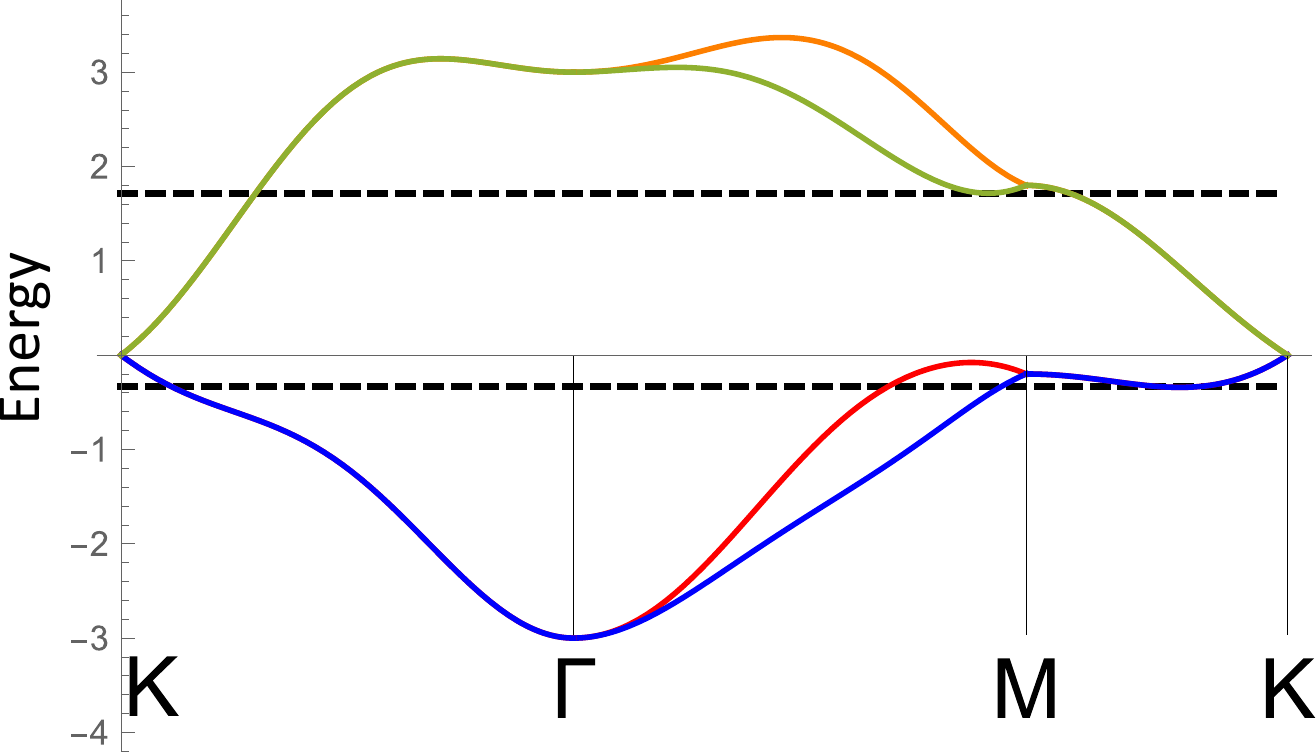} }
\centering{}\caption{Twisted bilayer graphene band structure calculated for $t_1 = 1, t_2 = -0.1, t_3 = 0.08$. The dashed lines show the positions of Van Hove singularities for electron-doping ($\mu \simeq 1.716  $) and hole-doping ($\mu \simeq -0.329 $). Note that the values of $\mu$ are different in the two cases. }
\label{fig_bands}
\end{figure}

In Fig. \ref{fig_bands} we show the calculated band structure for a particular choice of hopping magnitudes. This band structure is in good agreement with previously published results \cite{Bistritzer2011,Moon2012,Nam2017,Yuan2018,Kang2018PRX}.
In particular, it reproduces the splitting of the bands
along the $\Gamma M$-line, obtained in first-principles calculations \cite{Trambly2012,Fang2016,Cao2016PRL,Cao2018correlated} and effective low-energy models \cite{Bistritzer2011,Moon2012,Nam2017}. The bands are orbitally-polarized in terms of chiral orbitals $c_{\pm} = (c_x \pm i c_y)/ \sqrt{2}$.

\begin{figure}[h]
\begin{minipage}[h]{0.9\linewidth}
\center{\includegraphics[width=0.99\linewidth]{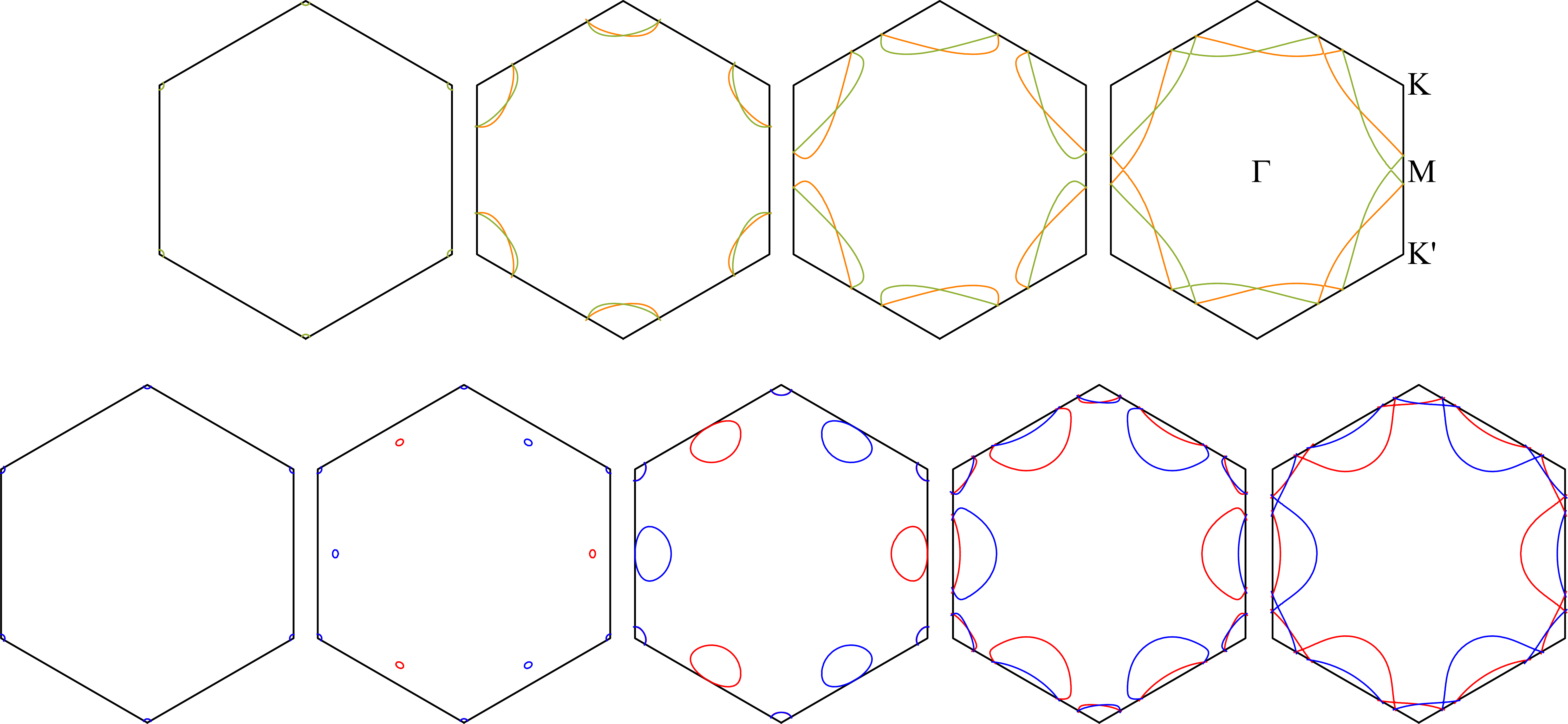}}
\end{minipage}
\centering{}\caption{Evolution of the Fermi surfaces with doping for the same hopping parameters as in Fig. \ref{fig_bands}. Top: evolution upon electron doping from charge neutrality (Dirac) point on the left
to Van Hove doping on the right. Bottom: same evolution  upon hole doping.}
\label{FSevol}
\end{figure}

For our purposes, the key feature of the band structure of Eq. (\ref{band_spec}) and  Fig. \ref{fig_bands} is that it allows for Lifshitz transitions at both positive and negative energies, and, hence, contains Van Hove points.
Some Lifshitz transitions lead to the appearance of VH points without logarithmically divergent density of states (DOS). For instance, decreasing $\mu$ down from the charge neutrality point $\mu =0$, one first reaches the Lifshitz transition at which isolated hole pockets centered on the $\Gamma M$ line appear (see Fig. \ref{FSevol}).
At such a transition, there is no VH  singularity in the DOS.  The reason is that VH points in this case are not saddle points, but local maxima of the band spectrum.
However, decreasing $\mu$ further, one reaches the value of $\mu$ at which another  Lifshitz transition occurs. This time, the corresponding VH points are saddle points of the dispersion, and  the DOS is logarithmically singular (this is what we earlier called a Van Hove singularity).
   The dashed lines in Fig. \ref{fig_bands} mark the values of chemical potential at which saddle-type VH points are located on the Fermi surface.   We will focus on these points because the large DOS increases the tendency towards superconductivity and competing orders.
 As we will be interested only in the states near the saddle-type VH points, we avoid a subtle issue
 whether in the presence of all symmetries of TBG,   the tight-binding model of Eq.~\eqref{band_spec}, based on localized Wannier states exclusively for the isolated flat bands (Refs. \cite{Yuan2018,Koshino2018PRX,Kang2018PRX}), is adequate everywhere in the Brillouin zone, or if there exist special $k-$points away from VH regions, where one needs to invoke other bands  to properly describe excitations \cite{PhysRevB.99.195455}.

\begin{figure}[h]
\begin{minipage}[h]{0.5\linewidth}
\center{\includegraphics[width=0.99\linewidth]{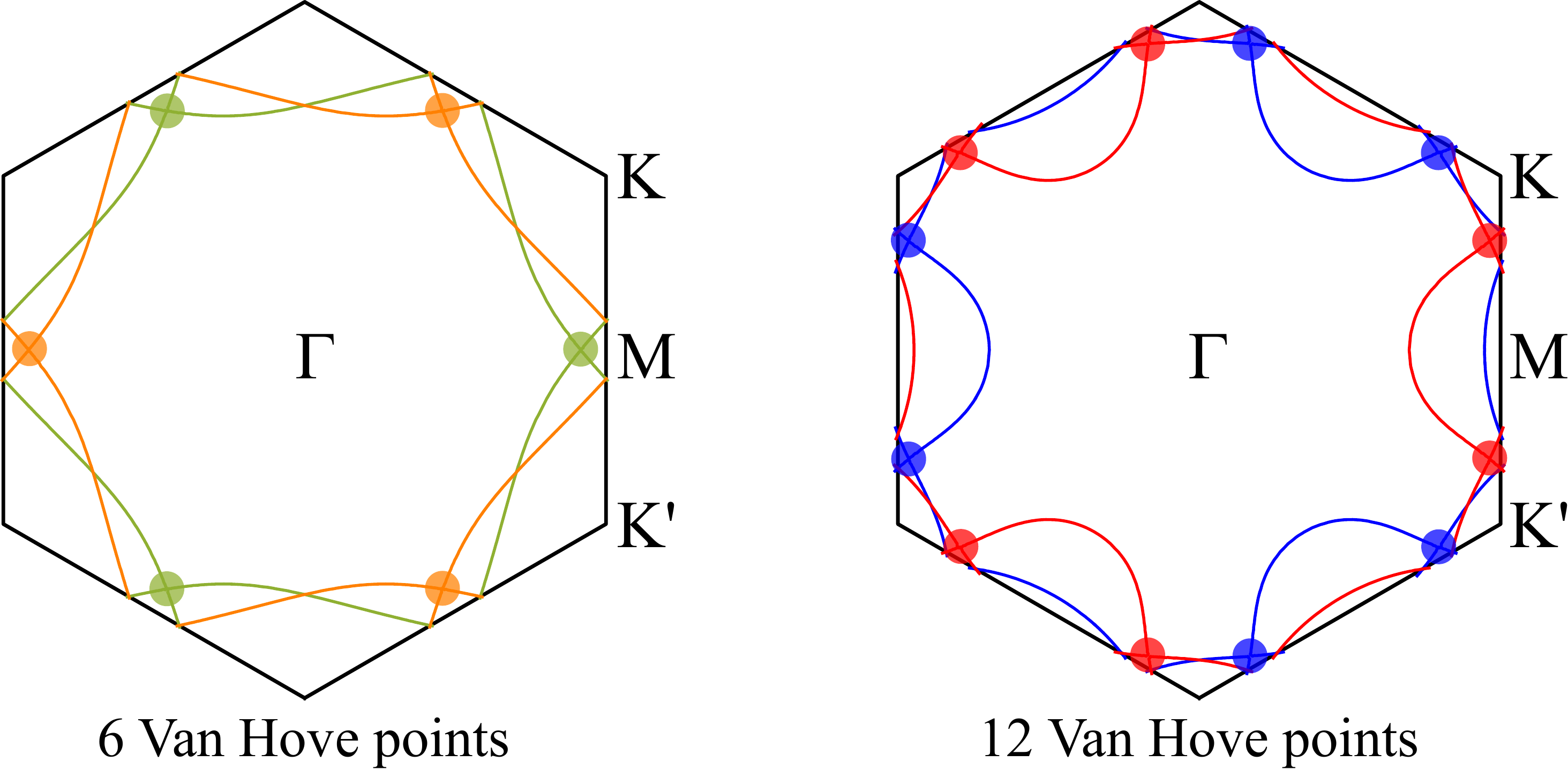}}
\end{minipage}
\centering{}\caption{Fermi surfaces for the  two bands (shown by different colors) for $\mu = 1.716$ (left) and $\mu = -0.329$ (right).
  Circles indicate the positions of Van Hove singularities. For  $\mu \simeq 1.716$ (electron doping) there are six
  Van Hove singularities, located along $\Gamma M$ and symmetry-related directions in the Brillouin zone. For $\mu \simeq -0.329$ (hole doping) there are twelve VH singularities; neither is located along a high-symmetry direction. }
\label{FStypes}
\end{figure}

In Fig. \ref{FSevol} we show the evolution of the Fermi surface for either hole or electron doping (negative or positive $\mu$).
We see different behavior in the two cases.
Upon electron doping, the system reaches a  Lifshitz transition with six VH
singularities,
located away from the Brillouin zone boundary, but along high symmetry $\Gamma M$ directions.
 Upon hole doping,  the first Lifshitz transition  creates additional pockets along the $\Gamma M$ lines, but  does not give rise to  VH singularities.
As $\mu$ decreases further,  the system does undergo a Lifshitz transition accompanied by VH singularities.  In this last case, there are twelve VH points in the Brillouin zone, each located away from the Brillouin zone boundary and also away from high symmetry directions.
  We show the Fermi surfaces with six  and twelve Van Hove singularities  separately in Fig. \ref{FStypes}.

For other values of hopping integrals we found three other scenarios: (a) Lifshitz transitions with six
VH singularities
for both electron and hole doping, (b) six
VH singularities at a Lifshitz transition for  hole doping and twelve for electron doping, and (ii)  Lifshitz transitions with twelve
VH singularities
for both electron and hole doping. In our analysis  below we focus on the Fermi-surface geometry in Figs.~\ref{fig_bands}, \ref{FStypes} , and \ref{FSevol},
because it appropriately describes the observed electron-hole asymmetry of the superconducting states in TBG~\cite{Cao2018unconventional,Cao2018correlated,Yankowitz2019,Lu2019}.
We also note in passing that as the twelve
VH singularities at the Lifshitz transition upon hole doping  form six sets of pairs  with small separation within a pair, there is the intriguing possibility~\cite{Yuan2019magic}  that
for fine-tuned hopping parameters,
VH points within each pair merge and create a set of six VH singularities, each leading to a stronger (power-law) divergence of the DOS~\cite{Isobe2019supermetal}.  We, however, will not study this special case.

\subsection{An effective low-energy patch model}
\label{sec:patch}

\begin{figure}[t]
\includegraphics[width=0.5\linewidth]{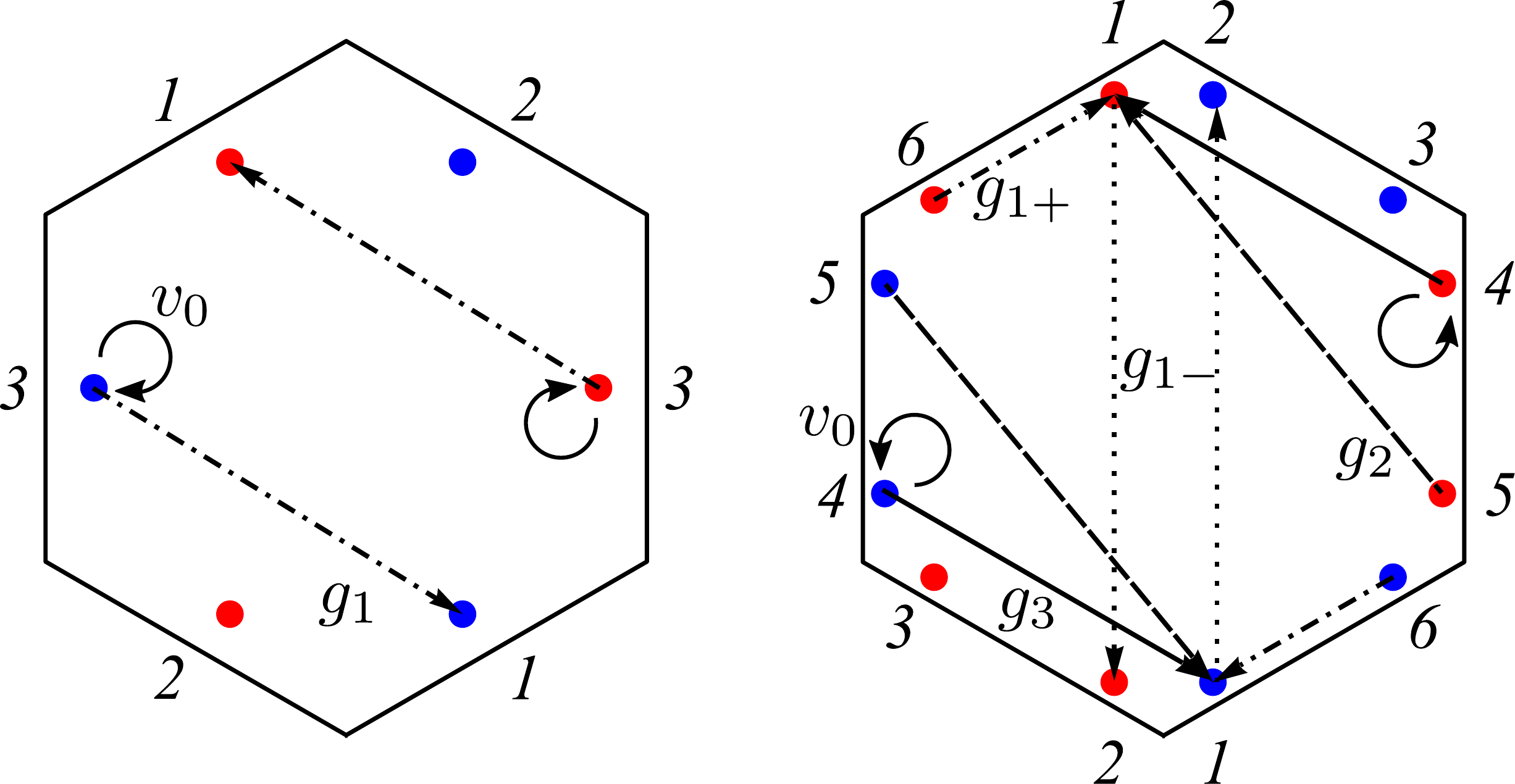}
\caption{ Sketch of the interactions in six-patch and twelve-patch models.
 Blue and red dots mark patches around Van Hove points.  We show only the interactions relevant for superconductivity.}
\label{mod_sk}
\end{figure}

Below we analyze superconductivity near Lifshitz transitions  accompanied by singularities in the DOS.
 For this we focus on states near the Van Hove points and introduce effective patch models with either six or twelve patches.
We first expand the energies Eq. (\ref{band_spec}) around the VH points and approximate the hopping Hamiltonian by
\begin{align}
\label{eq:Hpatch0}
H=\sum_{a=1}^{N_p} \sum_{\tau=\pm}\sum_{\sigma=\uparrow,\downarrow} \epsilon_{a}(\vec k) f_{a\tau\sigma}^\dagger(\vec k) f_{a\tau\sigma}(\vec k)\,,
\end{align}
where $f_{a\tau\sigma}(\vec k)$ annihilates an electron with momentum $\vec k$ in the vicinity of patch $a$ in band $\tau$ with spin $\sigma$. The patch index runs through $a=1\ldots N_p$ with $N_p=3$ ($6$) for the six-patch (twelve-patch) model.
The hyperbolic dispersion relations for the different patch points $\epsilon_{a}(\vec k)=\alpha_a k_x^2 - \beta_a k_y^2$ with sgn$(\alpha_a)=$ sgn$(\beta_a)$ are related by $D_3$ and time-reversal symmetry, inherited from the microscopic Hamiltonian of  Eq.~\eqref{2H}. Thereby, half of the patch points belong to one of the two bands crossing the Fermi energy at VH doping, while the other half belongs to the other band.
The patch points belonging to the same band are related by a threefold rotation symmetry, while the patches from different microscopic bands are related by inversion.
 We show the location of patches  in Figs. \ref{FStypes}, \ref{FSevol}, and \ref{mod_sk}.

We next consider all symmetry-allowed coupling terms between fermions in the patches.
In general, there are four types of allowed interactions. These are intra-patch and inter-patch density-density and exchange interactions. Umklapp processes are not allowed because VH singularities do not appear at
  momenta connected by a reciprocal lattice vector.  A simple bookkeeping analysis shows that there are 6 (18) symmetry-allowed couplings for the six-patch (twelve-patch) model without orbital-mixing terms, and 9 (27), when these terms are included.
 The orbital mixing terms were found to be very small numerically in the microscopic model~\cite{Kang2018strong,Koshino2018PRX}, and we do not include them.
The most general interacting Hamiltonian for the six-patch model  without orbital mixing is~~\cite{Isobe2018PRX}
\begin{align}
H^{Int}_{6p}= \sum_{a=1}^3\sum_{\tau=\pm}&\left[  u_0 f_{a\tau}^\dagger f_{a\tau} f_{a\tau}^\dagger f_{a\tau} +  \frac{v_0}{2} f_{a\tau}^\dagger f_{a\tau} f_{a\bar\tau}^\dagger f_{a\bar\tau} + u_1 f_{a\tau}^\dagger f_{a\tau} f_{a+1\tau}^\dagger f_{a+1\tau} +  v_1 f_{a\tau}^\dagger f_{a\tau} f_{a+1\bar\tau}^\dagger f_{a+1\bar\tau} + j_1 f_{a\tau}^\dagger f_{a+1\tau} f_{a+1\tau}^\dagger f_{a\tau} \right. \notag \\
&\left. +\left ( \frac{g_1}{2} f_{a\tau}^\dagger f_{a+1\tau} f_{a\bar\tau}^\dagger f_{a+1\bar\tau}+ \text{h.c.}\right)\right]
\label{eq:iap6}
\end{align}
 We introduced $\bar\tau=-\tau$ and $a$ labels half of the patches.
 We  omitted spin  indices for simplicity -- the spin structure of  each  term is $\sum_{\sigma,\sigma'}f_{\sigma}^\dagger f_\sigma f_{\sigma'}^\dagger f_{\sigma'}$.

For the twelve-patch model, the most general interaction Hamiltonian is
\begin{align}
H^{int}_{12p}= \sum_{a=1}^6\sum_{\tau=\pm}&\left[ u_0 f_{a\tau}^\dagger f_{a\tau} f_{a\tau}^\dagger f_{a\tau} +  \frac{v_0}{2} f_{a\tau}^\dagger f_{a\tau} f_{a\bar\tau}^\dagger f_{a\bar\tau} + u_2 f_{a\tau}^\dagger f_{a\tau} f_{a+2\tau}^\dagger f_{a+2\tau} +  v_2 f_{a\tau}^\dagger f_{a\tau} f_{a+2\bar\tau}^\dagger f_{a+2\bar\tau} + u_3 f_{a\tau}^\dagger f_{a\tau} f_{a+3\tau}^\dagger f_{a+3\tau} +  v_3 f_{a\tau}^\dagger f_{a\tau} f_{a+3\bar\tau}^\dagger f_{a+3\bar\tau} \right. \notag \\
& \left. + j_2 f_{a\tau}^\dagger f_{a+2\tau} f_{a+2\tau}^\dagger f_{a\tau} + \frac{g_2}{2} \left( f_{a\tau}^\dagger f_{a+2\tau} f_{a\bar\tau}^\dagger f_{a+2\bar\tau} + \text{h.c.} \right)+ j_3 f_{a\tau}^\dagger f_{a+3\tau} f_{a+3\tau}^\dagger f_{a\tau} + \frac{g_3}{2} \left( f_{a\tau}^\dagger f_{a+3\tau} f_{a\bar\tau}^\dagger f_{a+3\bar\tau} + \text{h.c.} \right) \right. \notag \\
&\left.  + u_{1+} f_{a\tau}^\dagger f_{a\tau} f_{a+(-1)^a\tau}^\dagger f_{a+(-1)^a\tau} + u_{1-} f_{a\tau}^\dagger f_{a\tau} f_{a-(-1)^a\tau}^\dagger f_{a-(-1)^a\tau}  + v_{1+} f_{a\tau}^\dagger f_{a\tau} f_{a+(-1)^a\bar\tau}^\dagger f_{a+(-1)^a\bar\tau} + v_{1-} f_{a\tau}^\dagger f_{a\tau} f_{a-(-1)^a\bar\tau}^\dagger f_{a-(-1)^a\bar\tau}  \right. \notag \\
& + j_{1+} f_{a\tau}^\dagger f_{a+(-1)^a\tau} f_{a+(-1)^a\tau}^\dagger f_{a\tau} + \frac{g_{1+}}{2}  \left( f_{a\tau}^\dagger f_{a+(-1)^a\tau} f_{a\bar\tau}^\dagger f_{a+(-1)^a\bar\tau} + \text{h.c.} \right) \notag \\
&\left. + j_{1-} f_{a\tau}^\dagger f_{a-(-1)^a\tau} f_{a-(-1)^a\tau}^\dagger f_{a\tau} + \frac{g_{1-}}{2} \left( f_{a\tau}^\dagger f_{a-(-1)^a\tau} f_{a\bar\tau}^\dagger f_{a-(-1)^a\bar\tau}+ \text{h.c.} \right) \right]\,
\label{eq:iap12}
\end{align}
with the patch index being defined modulo 6.

\begin{figure}[t]
\begin{minipage}[h]{0.2\linewidth}
\center{\includegraphics[width=0.99\linewidth]{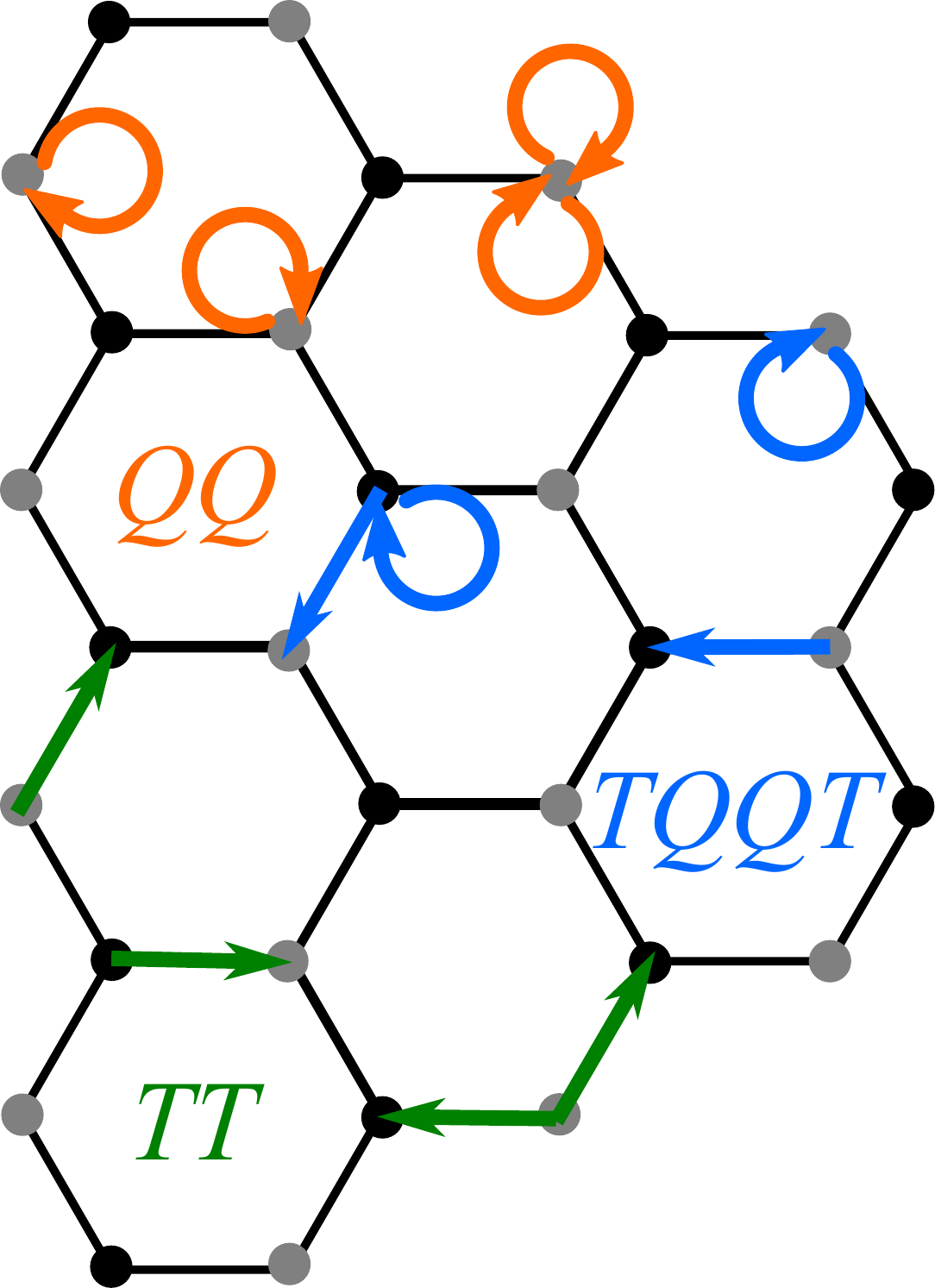}}
\end{minipage}
\centering{}\caption{Graphic representation of three types of interactions in the Hamiltonian of Eq. (\ref{int_rsp}). $QQ$ (orange) are density-density interactions, $TQ$ and $QT$ (blue) describe processes, in which an electron interacts with a local density while hopping to a neighboring site, and  $TT$ (green)
  describes pair hopping processes.}
\label{scat_pic}
\end{figure}

Below, we will need the subset of interactions relevant to pairing, which are between fermions with opposite momenta. In our model these fermions  belong to different bands. The pairing interactions  then only involve fermions with patch indices $(a,\tau)$ and $(a,\bar\tau)$ (see Fig. \ref{mod_sk}). This reduces the number of interaction terms relevant for SC to two, $v_0, g_1$, for the six-patch model and to five,  $v_0, g_2,g_3,g_{1+}, g_{1-}$, for the twelve-patch model.  We sketch the interactions which will be important for the pairing problem
  in Fig. \ref{mod_sk}.

To estimate the values of the couplings, we  need to compare Eqs. (\ref{eq:iap6}) and (\ref{eq:iap12}) with the corresponding interaction terms in the microscopic model.
A typical approximation for the four-fermion interaction term for a system with screened Coulomb interaction is to keep it local, i.e., approximate the interaction by the on-site Hubbard density-density interaction.
 The case of TBG was argued to be different,  because there is  substantial overlap between Wannier states  localized at neighboring sites \cite{Koshino2018PRX,Kang2018strong}. This peculiar property leads to a new form of the interaction Hamiltonian \cite{Kang2018strong,Koshino2018PRX}, in which local density-density interactions and terms describing assisted nearest neighbor hopping are of the same order and have to be considered on equal grounds.
We follow Kang and Vafek \cite{Kang2018strong} and write the interaction Hamiltonian in real space as
\begin{equation}
H_{int} =
V_0 \sum_{\bold{R}} \left( \sum_{o=x,y} \sum_{\sigma=\uparrow,\downarrow} O_{o,\sigma} (\bold{R}) \right)^2,
\label{int_rsp}
\end{equation}
where
\begin{align}
O_{o, \sigma} (\bold{R}) &= \frac{1}{3} Q_{o,\sigma} (\bold{R}) + \alpha_T T_{o,\sigma} (\bold{R}), \\
Q_{o,\sigma} (\bold{R}) &= \sum_{n=1}^3 \left( c^{\dagger}_{o \sigma A} (\bold{R} + \vec{a}_n) c_{o \sigma A} (\bold{R} + \vec{a}_n) + c^{\dagger}_{o \sigma B} (\bold{R} - \vec{a}_n) c_{o \sigma B} (\bold{R} - \vec{a}_n) \right), \\
T_{o,\sigma} (\bold{R}) &= \sum_{n=1}^3 \left( c^{\dagger}_{o \sigma B} (\bold{R} - \vec{a}_n) c_{o \sigma A} (\bold{R} + \vec{a}_n) - c^{\dagger}_{o \sigma A} (\bold{R} + \vec{a}_{n+1}) c_{o \sigma B} (\bold{R} - \vec{a}_n) + h.c. \right).
\end{align}
Here, $\bold{R}$ runs over the centers of the honeycomb lattice corresponding to the triangular moir\'e pattern, $A, B$ denote the sublattice indexes, $ \vec{a}_n$ are three translation vectors of the honeycomb sites (see Fig.~\ref{lattice}) and $o$ is the Wannier orbital index, inherited from the original valley degrees of freedom.
There are  three types of interaction terms:  $QQ$, $TQ, QT$, and $TT$ terms. The $QQ$ term describes local  (Hubbard)  density-density interactions  within a single honeycomb,
the $TQ,QT$ terms describe the processes, in which an electron interacts with a local density while hopping to a neighboring site, and the  $TT$ term describes the pair-hopping  processes, in which  electrons interact with each other, while hopping to a neighboring site
  (see Fig. \ref{scat_pic}).
In momentum space, $H_{int}$ becomes
\begin{align}
H_{int}= \sum_{k,q,k',q'} \sum_{\sigma,\sigma',o,o'} \delta(k-q+k'-q') \biggr[ &
\frac{1}{9} \sum_{v,v'} F_{vv'v'v} c^{\dagger}_{o \sigma v k}  c^{\dagger}_{o' \sigma' v' k'} c_{o' \sigma' v' q'} c_{o \sigma v q} \notag \\
& + \frac{\alpha_T}{3} \sum_{v \neq v'} \biggr( F_{vvvv'} c^{\dagger}_{o \sigma v k}  c^{\dagger}_{o' \sigma' v k'} c_{o' \sigma' v q'} c_{o \sigma v' q} + F_{vvv'v} c^{\dagger}_{o \sigma v k}  c^{\dagger}_{o' \sigma' v k'} c_{o' \sigma' v' q'} c_{o \sigma v q} \notag \\
&\qquad \quad+ F_{vv'vv} c^{\dagger}_{o \sigma v k}  c^{\dagger}_{o' \sigma' v' k'} c_{o' \sigma' v q'} c_{o \sigma v q} +  F_{v'vvv} c^{\dagger}_{o \sigma v' k}  c^{\dagger}_{o' \sigma' v k'} c_{o' \sigma' v q'} c_{o \sigma v q} \biggr) \notag \\
&+ \alpha_T^2 \sum_{v \neq v'} \left( F_{vv'vv'} c^{\dagger}_{o \sigma v k}  c^{\dagger}_{o' \sigma' v' k'} c_{o' \sigma' v q'} c_{o \sigma v' q} + F_{vvv'v'} c^{\dagger}_{o \sigma v k}  c^{\dagger}_{o' \sigma' v k'} c_{o' \sigma' v' q'} c_{o \sigma v' q} \right)
 \biggr],
\label{inter_k}
\end{align}
where $v,v'$ label the sublattice indexes $A$ and $B$, $F_{vv'v''v'''}$ are coupling functions, which we present in Supplementary material, and $\alpha_T$ measures the strength of non-local interactions
 (the $QQ$, $TQ, QT$, and $TT$ terms are
$O(1), O(\alpha_T)$ and $O(\alpha^2_T)$ terms, respectively).  Kang and Vafek estimated $\alpha_T$ to be around
$1/4$.
  We will use $\alpha_T$ as a parameter, but keep it close to $1/4$.

We next transform  $H_{int}$  to the band basis and project it onto the patches around VH points. We show the details in the Supplementary material  and here present the results.
 For the interactions relevant to SC, we obtain,
in units of $V_0$ from Eq.~(\ref{int_rsp}),
\begin{equation}
v_0 = 1      \quad  g_1 = 0.1 +0.92 \alpha^2_T
\label{tu_1}
\end{equation}
for the six-patch model and
\begin{align}
v_0 &=1      \quad g_2 = 0.193 + 0.053 \alpha^2_T \quad g_{3} = 0.021 + 1.51 \alpha^2_T  \notag \\
 g_{1+}&= 0.256 + 17.9 \alpha^2_T \quad  g_{1-} = 0.057 + 9.02 \alpha^2_T
\label{yu_2}
\end{align}
for the twelve-patch model. Observe  that the prefactors of the $\alpha^2_T$ terms are large numbers.

 The interactions in Eqs. (\ref{tu_1}) and (\ref{yu_2}) are the bare ones. The true
  interactions  relevant to superconductivity are  the effective, fully irreducible ones, which includes all corrections
 from particle-hole bubbles  and also all renormalizations in the particle-particle channel from fermions with energies above the characteristic scale, which is, roughly, the largest of $T_c$ and the Fermi energy at the VH points.
  The flow of the couplings upon integrating out fermions with higher energies is often described within renormalization group (RG) approach~\cite{MaitiSCrepuls,RevModPhys.66.129,RevModPhys.84.299,FRGreview2}.
  These renormalizations are particularly relevant when the bare interaction is repulsive in all pairing channels, as renormalizations may overcome the bare repulsion and make the  interaction attractive
   in one or more pairing channels, below certain energies.   Physically, these renormalizations make the effective interaction non-local, and the growing non-local component eventually  gives rise to a sign change of the pairing interaction in certain channels.
    In our case, the bare interaction is already non-local, and we show in the next section  that it is already attractive in one channel for the six-patch model and in two channels for the twelve-patch model.  In this situation, the RG-type  renormalization of the bare interaction may affect the magnitudes of the attractive couplings, but will unlikely change qualitatively the results obtained with the bare interactions. We therefore proceed without including the RG flow of the couplings.  We emphasize that here we only focus on the pairing channel and do not address the issue of competing orders. To study the interplay between superconductivity and competing orders, RG-type calculations are required.

We also note in passing that previous works did apply RG to both six-patch models~\cite{Lin2018,Lin2019,Isobe2018PRX} and a twelve-patch model~ \cite{Gonzalez2019} for TBG. However, these
 works considered the cases when the RG flow of the couplings (or, at least, Kohn-Luttinger renormalizations from particle-hole bubbles) is necessary to overcome a bare repulsion and induce an attractive pairing interaction.

\section{Gap equation}
\label{sec:gap}

\begin{figure}[t]
\center{\includegraphics[width=0.5\linewidth]{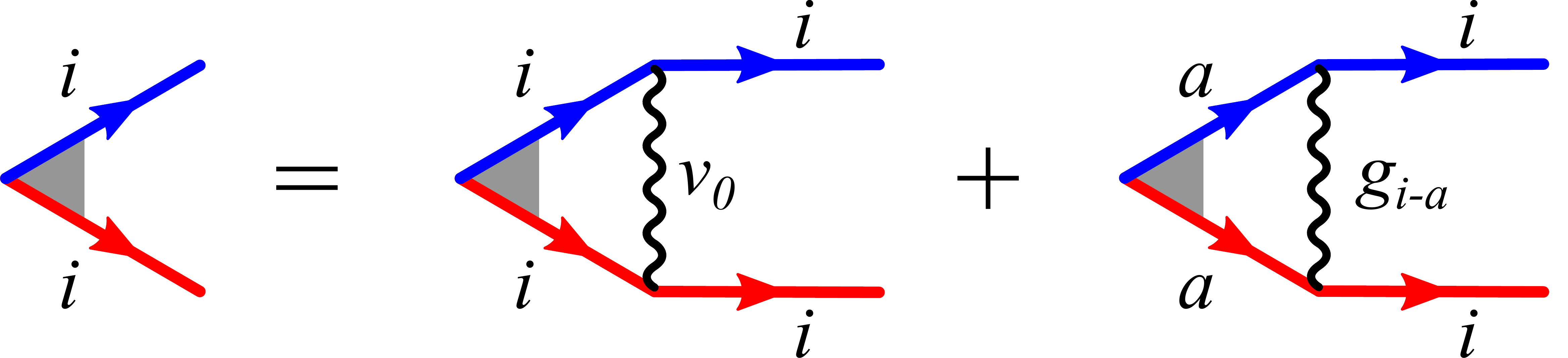} }
\centering{}\caption{Diagrammatic representation of a system of coupled gap equations. Gray triangle is a fully renormalized superconducting vertex, red and blue lines are Green's functions of  electrons from the two bands. Summation over $a$ is implied.}
\label{gapeq_pic}
\end{figure}

To study superconductivity,
 we introduce  the gap function as $\Delta_a^{\sigma\tau\sigma'\tau'}=\langle f_{a\tau\sigma} f_{a\tau'\sigma'} \rangle$, where, we remind, $a$ labels the  patches, and $\tau$  and $\sigma$ are band and spin indices.
 In the absence of orbital mixing,  spin-singlet and spin-triplet channels are degenerate in TBG  because direct exchange between  patches related by time inversion is absent~\cite{Isobe2018PRX}.
  A finite orbital mixing slits  spin-singlet and triplet channels.
Depending on the sign of the orbital mixing term, either spin-singlet or spin-triplet  SC will be favored \cite{Isobe2018PRX,You2019npj}.
Experimentally, superconductivity in TBG is destroyed by small magnetic fields \cite{Cao2018unconventional}, which is consistent with spin-singlet pairing.  We therefore will  focus on spin-singlet pairing.
  We assume that the orbital mixing term is smaller than the other interaction terms. In this situation, the Cooper  pairs
  with zero total momentum
  are predominantly made by fermions from different bands, see Figs.~\ref{mod_sk} and \ref{gapeq_pic}.  Accordingly, wee set  $\tau' = {\bar \tau}$ in
  $\Delta_a^{\sigma\tau\sigma'\tau'}$ and express it as $\Delta_a^{\sigma\tau\sigma'\bar\tau}=\Delta_a i \sigma_y$ ($\Delta_a$ is the same for the two choices of $\tau$).
The matrix gap equation then reduces to a set of three (six) coupled equations for the six-patch (twelve-patch) model:
\begin{align}
\vec{\Delta}_{6p} &= -\Pi \begin{pmatrix}
v_0 & g_1 & g_1  \\
g_1 & v_0 & g_1  \\
g_1 & g_1 & v_0
\end{pmatrix}\vec{\Delta}_{6p},\\
\vec{\Delta}_{12p} &= -\Pi \begin{pmatrix}
v_0 & g_{1-} & g_2 & g_3 & g_2 & g_{1+} \\
g_{1-} & v_0 & g_{1+} &  g_2 & g_3 & g_2  \\
g_2 & g_{1+} & v_0 & g_{1-} & g_2 & g_3 \\
g_3 & g_2 & g_{1-} & v_0 & g_{1+} & g_2 \\
g_2 & g_3 & g_2 & g_{1+} & v_0 & g_{1-} \\
g_{1+} & g_2 & g_3 & g_2 & g_{1-} & v_0
\end{pmatrix}
\vec{\Delta}_{12p}.
\label{gapeq}
\end{align}
Here $\vec{\Delta}_{6p}=(\Delta, \Delta_2,\Delta_3)^T$, $\vec{\Delta}_{12p}=(\Delta, \Delta_2,\Delta_3,\Delta_4,\Delta_5,\Delta_6)^T$, and $\Pi$ is the particle-particle polarization bubble (the same for all pairs of fermions).
Diagonalizing the matrix gap equation, we obtain  eigenvalues and eigenfunctions in different pairing channels.
  We classify the solutions of the gap equation according to the irreducible representations of the point group
   $D_3 = C_3 \times C_2$, whose elements are rotations
  along the z-axis by $\pm2\pi/3$ ($C_3$) and twofold rotations along the y- and symmetry-equivalent axes ($C_2$), Ref. ~\cite{Hamermesh}.
 The $D_3$ group has two one-dimensional irreducible representations, called $A_1$ and $A_2$, and one  two-dimensional  representation, called $E$ (the corresponding eigenvalue is doubly degenerate).
 Each representation contains an infinite set of eigenfunctions, some describe spin-singlet and some spin-triplet order.  The generic form of eigenfunctions in $A_1$ is $\cos (6n\theta_k)$ for spin-singlet pairing ($n =0,1,2..$) and  $\sin((6n+3) \theta_k)$ for spin-triplet pairing with the polar angle $\theta_k=\arctan k_y/k_x$ counted from the $k_x$ axis.
 For $A_2$, the eigenfunctions are $\cos((6n+3) \theta_k)$ for spin-triplet
 and $\sin(6 n \theta_k)$ for spin-singlet
 pairing.
 For $E$, the eigenfunctions  are $\left(\cos((6n+2) \theta_k),\sin((6n+2) \theta_k)\right)$ and $\left(\cos((6n +4) \theta_k),\sin((6n +4) \theta_k)\right)$ for spin-singlet pairing and  $\left(\cos((6n+1) \theta_k),\sin((6n+1) \theta_k)\right)$ and $\left(\cos((6n+5) \theta_k),\sin((6n+5) \theta_k)\right)$ for spin-triplet pairing.  The gap equation decouples between different representations, but not between different eigenfunctions within the same representation.
  In a generic case, when all Fermi surface points are relevant to pairing,  all partial components get coupled in the gap equation.  In patch models, the  gap equation simplifies because only a limited number of harmonics is distinguishable.
  For simplicity, we will use the lowest harmonics to describe our solution of the gap equation.

 In our sign convention, a specific channel becomes attractive when the eigenvalue turns from negative to positive. We show below that if we keep only local $QQ$ terms in the interaction (i.e., set $\alpha_T=0$),  all eigenvalues are negative and superconductivity does not occur without additional contributions to the pairing interaction from, e.g., Kohn-Luttinger diagrams.  However, once we add non-local terms,
  we find that some channels become attractive once $\alpha_T$ exceeds some critical value, specific to a given channel.

Solving the gap equation for the six-patch model, we find that only one eigenfunction from $A_1$ and one from $E$ contribute to  spin-singlet pairing.
To express the corresponding eigenfunctions, we note that
the patches are centered along high symmetry directions. In this situation,
the polar angles of the patch locations $\theta_i$,
$i=1,2,3$, are related by $\theta_2 = \theta_1 - \pi/3$, and $\theta_3 = \theta_1 - 2\pi/3$.
We can then write the eigenfunctions  as
 \bea
{\vec \Delta}^{A_1}_{6p} &=& (1,1,1)  \nonumber \\
{\vec \Delta}^{E_1}_{6p} &=& \left(\cos(2 \theta_1),\cos(2 \theta_1 - 2 \pi/3),\cos(2 \theta_1 + 2 \pi/3)\right) \nonumber \\
{\vec \Delta}^{E_2}_{6p} &=& \left(\sin(2 \theta_1),\sin(2 \theta_1 - 2 \pi/3),\sin(2 \theta_1 + 2 \pi/3)\right).
\label{wed_1}
\eea
  The eigenfunction in the $A_{1}$ representation  has the same sign in all patches and is analogous to an $s-$wave.
  The eigenfunctions in the $E$ representation  change sign four times as one makes the full circle along the  Fermi surface, and in this respect are analogous  to $d-$wave.

\begin{figure}[t]
\includegraphics[width=0.99\linewidth]{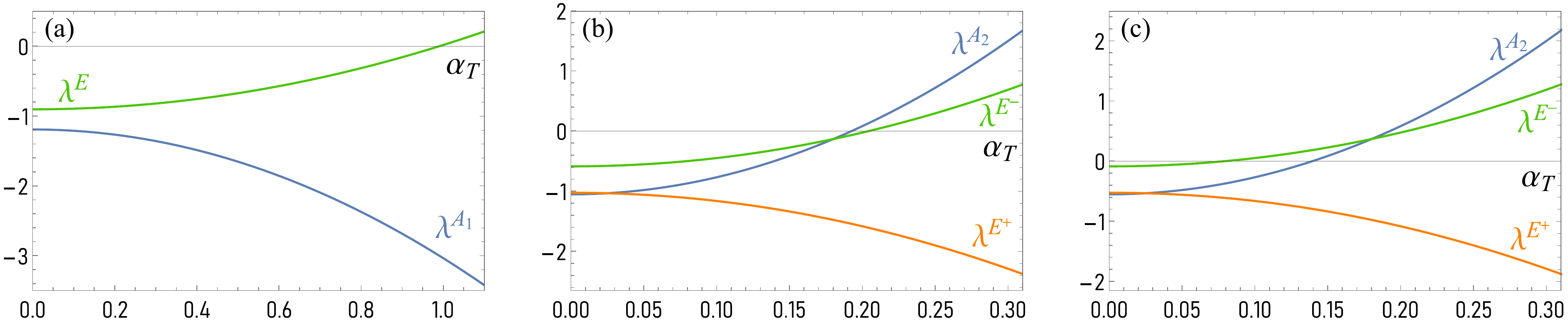}
\centering{}\caption{The eigenvalues for the six-patch model $\lambda^E$ and $\lambda^{A_1}$ (a) and for the twelve-patch model, $\lambda^{E^+}, \lambda^{E^-}$,  and $\lambda^{A_2}$ ((b) and (c)), as functions of $\alpha_T$ ($\lambda^{A_1}$ is irrelevant and not shown).
  When an eigenvalue turns positive, the interaction in the corresponding pairing channel becomes attractive. For the  six patch model, $\lambda^E >0$ for $\alpha_T >0.98$. For the twelve patch model, $\lambda^{E^-}$,  and $\lambda^{A_2}$ become positive when $\alpha_T$ exceeds certain values. Panel (b) is for the interactions, extracted from the microscopic model (see text). For panel (c), we increased exchange interactions by a factor of two. Observe that $\lambda^{E^-}$ and $\lambda^{A_2}$ are nearly degenerate over some range of $\alpha_T$. }
\label{la}
\end{figure}

The eigenvalues in the $A_1$ and $E$ channels are
\begin{align}
\label{eigvals6}
\lambda_{6p}^{A_1}&=- \Pi(v_0+2g_1)\\
\lambda_{6p}^{E}&=- \Pi(v_0-g_1).
\end{align}
Substituting the values of couplings from (\ref{tu_1}), we obtain
\begin{align}
\label{eigvals6_1}
\lambda_{6p}^{A_1}&=- \Pi V_0 (1.2+1.84\alpha_T^2)\\
\lambda_{6p}^{E}&=- \Pi V_0 (0.9-0.92\alpha_T^2).
\end{align}
We plot the eigenvalues as functions of $\alpha_T$  in the left panel of Fig.~\ref{la}.
 We see that the coupling in the $A_1$ channel is negative (repulsive) for all values of $\alpha_T$,  but the one in the $E$ channel becomes attractive for $\alpha_T > 0.98$.

\begin{figure}[t]
\center{\includegraphics[width=0.5\linewidth]{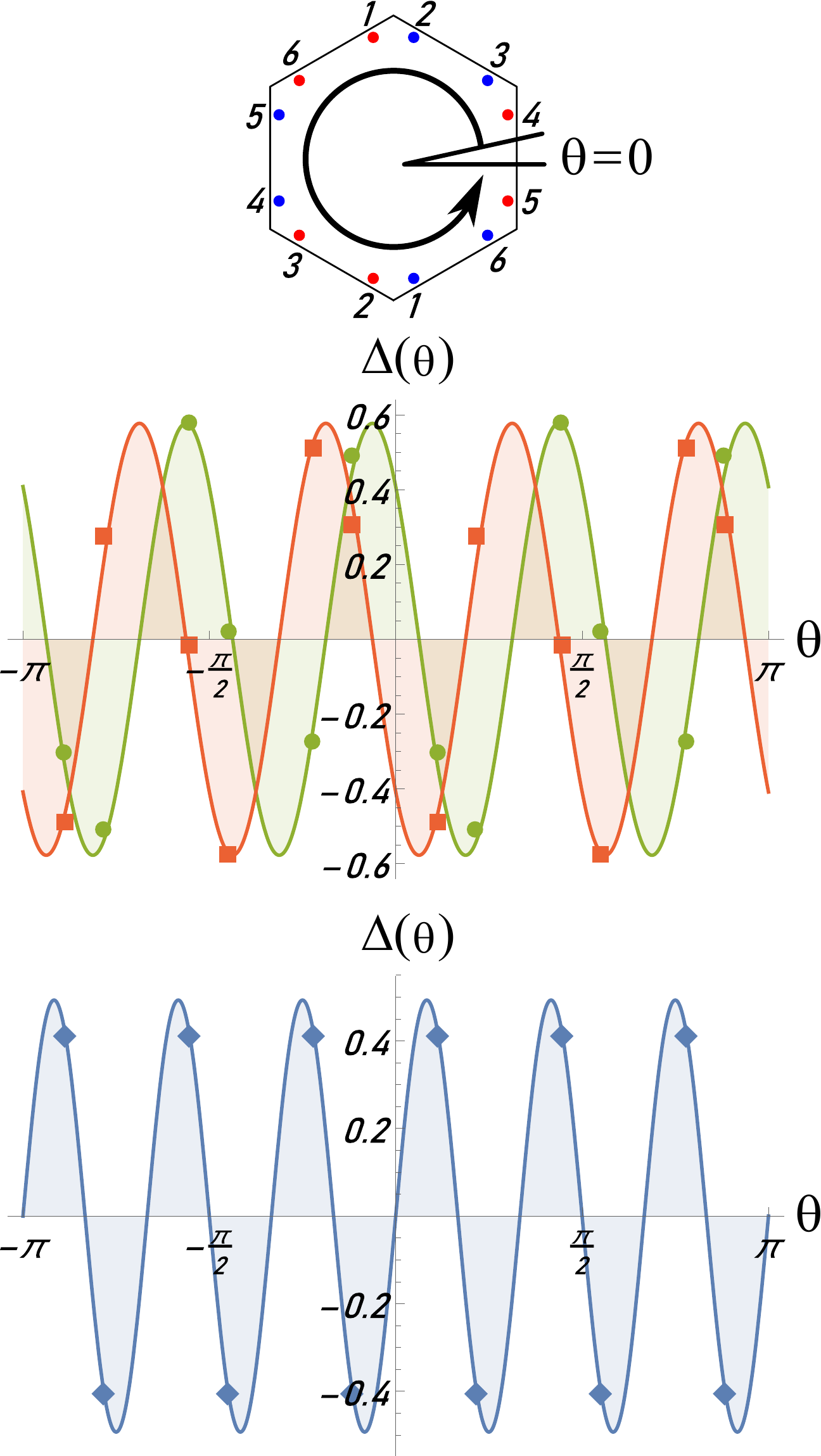} }
\centering{}\caption{The eigenfunctions  for $E$ and $A_2$ states in the twelve-patch model.
Numbering of Van Hove points is shown on top.
 Middle:  components
${\vec \Delta}_{12p}^{E_1} =  \cos(4\theta_i + 3\pi/4)$ (red) and ${\vec \Delta}_{12p}^{E_2} =
\sin(4\theta_i + 3\pi/4)$ (green).
Bottom: ${\vec \Delta}_{12p}^{A_2} = \sin(6\theta_i)$ (blue).
 Circles, squares, and diamonds denote the  values of gap function at different VH points. Lines are continuous functions of $\theta$, obtained using symmetry  reasoning.
  Viewed as continuous functions,  ${\vec\Delta}_{12p}^{E_1}$ and  ${\vec\Delta}_{12p}^{E_2}$ have eight nodes,
  and ${\vec \Delta}_{12p}^{A_2}$ has twelve nodes.}
\label{3gapsol}
\end{figure}

 For the twelve-patch model,
  $\theta_{2m+1} =  \theta_1 - m\pi/3$, $\theta_2 = \pi - \theta_1$, and
   $\theta_{2m+2} = \theta_2 - m\pi/3$ (see Fig. \ref{3gapsol}).
    This doubles the number of non-equivalent eigenstates and eigenfunctions.
        The   eigenfunctions for spin-singlet pairing  are
   \bea
 {\vec \Delta}^{A_1}_{12p} &=& (1,1,1,1,1,1)  \nonumber \\
 {\vec \Delta}^{A_2}_{12p} &=& (-1,1,-1,1,-1,1) \nonumber \\
{\vec \Delta}^{E^{+}_1}_{12p} &=& \left(\cos(2 \theta_1),\cos(2 \theta_1),\cos(2 \theta_1 - 2\pi/3),\cos(2 \theta_1 + 2\pi/3),\cos(2 \theta_1+ 2\pi/3),\cos(2 \theta_1 - 2\pi/3)\right) \nonumber\\
{\vec \Delta}^{E^{+}_2}_{12p} &=& \left(\sin(2 \theta_1),-\sin(2 \theta_1),\sin(2 \theta_1 - 2\pi/3),-\sin(2 \theta_1 + 2\pi/3),\sin(2 \theta_1+ 2\pi/3),-\sin(2 \theta_1 - 2\pi/3)\right) \nonumber\\
{\vec \Delta}^{E^{-}_1}_{12p} &=& \left(\cos(4 \theta_1),\cos(4 \theta_1),\cos(4 \theta_1+2\pi/3),\cos(4 \theta_1-2\pi/3),\cos(4 \theta_1-2\pi/3),\cos(4 \theta_1+2\pi/3)\right) \nonumber\\
{\vec \Delta}^{E^{-}_2}_{12p} &=& \left(\sin(4 \theta_1),-\sin(4 \theta_1),\sin(4 \theta_1+2\pi/3),-\sin(4 \theta_1-2\pi/3),\sin(4 \theta_1-2\pi/3),-\sin(4 \theta_1+2\pi/3)\right) \nonumber\\
 \label{wed_2}
 \eea
  With respect to the number of nodes, the $A_1$ eigenfunction corresponds to $s-$wave. The $A_2$ eigenfunction is proportional to $\sin(6\theta_k)$ at the patch points and corresponds to $i$-wave (12 nodes along the Fermi surface). Eigenfunctions $E^+$ and $E^-$ from $E$ correspond to $d-$wave and $g-$wave, respectively  (4 nodes and 8 nodes,   see Fig. \ref{3gapsol}).
      Because the two states with $E$ symmetry have different number of nodes, they decouple in the gap equation (this does not hold beyond  the patch model).

The eigenvalues in the four decoupled channels are
\begin{align}
\lambda_{12p}^{A_{1/2}} &= - \Pi \left[ v_0 + 2 g_2 \pm (g_{1-} + g_{1+} + g_3) \right], \\
\lambda_{12p}^{E^\pm} &= -\Pi \left [v_0 - g_2 \pm \sqrt{g_{1-}^2 +g_{1+}^2 + g_3^2 - g_{1-} g_{1+}  - g_{1-} g_3 - g_{1+} g_3  } \right].
\label{eigvals}
\end{align}

Substituting the values for the couplings from (\ref{yu_2}), we obtain
\begin{align}
\lambda_{12p}^{A_{1/2}} &= - \Pi V_0\left[1.39 + 0.106 \alpha^2_T \pm (0.33 + 28.43 \alpha^2_T ) \right], \\
\lambda_{12p}^{E^\pm} &= -\Pi V_0\left [ 0.81 -0.053 \alpha^2_T \pm \sqrt{0.05 + 5.89 \alpha^2_T + 201.94 \alpha^4_T}\right].
\label{eigvals_1}
\end{align}
We plot the eigenvalues as functions of $\alpha_T$ in the right panel of Fig.~\ref{la}.
We see that $\lambda_{12p}^{A_{1}} $ and $\lambda_{12p}^{E^+}$ are repulsive for all values of $\alpha_T$, but $\lambda_{12p}^{A_{2}} $ and $\lambda_{12p}^{E^-}$  become attractive for $\alpha_T > 0.19$ and $0.21$, respectively.
 Note that the values of $\alpha_T$ needed for attraction are smaller than in the six-patch model and also smaller than the estimate for $\alpha_T\sim 0.23$
  presented in  Ref.~\cite{Kang2018strong}.
 Furthermore, over some range of $\alpha_T \sim 1/4$, the couplings $\lambda_{12p}^{A_{2}} $ and $\lambda_{12p}^{E^-}$ are almost identical, i.e., the critical temperatures $T^{A_2}_c$ and $T^{E}_c$ are approximately the same.

 We emphasize that this observation represents a qualitative difference to previous studies of superconductivity within patch models for Van Hove filling in TBG~\cite{Gonzalez2013,Lin2018,Gonzalez2019} because in our case
 no Kohn-Luttinger type corrections (or higher order corrections associated with spin or charge fluctuations) are needed to induce attractive pairing interactions.
 As a consequence, we anticipate a higher critical temperature than typically expected for Kohn-Luttinger type superconductivity.  Fluctuation corrections will increase non-local interactions, i.e. shift $\alpha_T$ to a larger value and somewhat increase $T^{A_2}_c$ and $T^{E}_c$ from  already non-zero values.

In the next section we derive the Landau free energy and  analyze the SC state below $T_c$. We show that the near-degeneracy between the eigenfunctions in the $A_2$ and $E^-$ channels  leads to a highly non-trivial phase diagram with a large region of the coexistence state,
 where superconductivity breaks not only the $U(1)$ phase symmetry, but also three-fold $C_3$ lattice rotational symmetry, i.e., the SC state is a nematic superconductor. We show that two types of nematic superconductors emerge, depending on system parameters. One additionally breaks time reversal symmetry,  the other preserves it.

 Before we proceed, we make an adjustment of the $E^-$ state in the 12-patch model for further convenience.
 Namely, we use the fact that  any rotation of
the two components of $E^-$ is still an eigenfunction and rotate them by $3\pi/4$.
 The new components $E^-_1 = E_1$ and $E^-_2 = E_2$, are
    \begin{align}
\vec{\Delta}_{12p}^{E_1}&=\big(\cos (4\theta_1+3\pi/4),\ldots ,\cos (4\theta_6 + 3\pi/4)\big)  \nonumber \\
\vec{\Delta}_{12p}^{E_2}&=\big(\sin (4\theta_1+3\pi/4),\ldots ,\sin (4\theta_6+3\pi/4)\big)
\label{wed_3}
\end{align}
With this choice of  basis eigenfunctions, $\Delta_{12p}^{E_1}$ and $\Delta_{12p}^{E_2}$ exchange as  $\Delta_{12p}^{E_{1,2}} \to -\Delta_{12p}^{E_{2,1}}$ under
the twofold rotations around $k_y$ and symmetry related axes.

\section{Landau free energy and the structure of the superconducting state}
\label{sec:GL}

We  express the gap function in the superconducting state as a linear combination of the eigenfunctions for the attractive pairing components. In the six-patch model we have
 \begin{equation}
\Delta^{SC}_{6p}=\Delta_{E_1} \vec\Delta_{6p}^{E_1} + \Delta_{E_2} \vec\Delta_{6p}^{E_2},
\end{equation}
 where $\Delta_{E_{1}}$ and $\Delta_{E_{2}}$ are complex numbers.
In the twelve-patch model we have
\begin{equation}
\Delta^{SC}_{12p}=\Delta_{A_2}\vec\Delta_{12p}^{A_2} + \Delta_{E_1}\vec\Delta_{12p}^{E_1} + \Delta_{E_2} \vec\Delta_{12p}^{E_2},
\label{eq:12pgap}
\end{equation}
where $\Delta_{E_{1}}$,  $\Delta_{E_{2}}$, and $\Delta_{A_2}$ are complex numbers.

To analyze superconducting ground states,
 we derive the Landau free energy,
 $\mathcal{F}_{6p} =  \mathcal{F}_{6p} (\Delta_{E_{1}}, \Delta_{E_{2}})$ and   $\mathcal{F}_{12p} =  \mathcal{F}_{12p} (\Delta_{E_{1}}, \Delta_{E_{2}}, \Delta_{A_2})$, find their minima and obtain
   the  magnitudes and phases of  $\Delta_{E_{1}}$ and $\Delta_{E_{2}}$ for the six-patch model and  of $\Delta_{A_2}$,  $\Delta_{E_{1}}$, and  $\Delta_{E_{2}}$  for the twelve-patch model.
The functional form of the Landau free energy for each model is  determined by $D_3$ and $U(1)$ symmetries \cite{RevModPhys.63.239}, however which superconducting state is realized depends on  the parameters of the Landau free energy.
 We obtain these parameters by applying a Hubbard-Stratonovich decomposition to the underlying fermionic model and integrating out fermions (see Supplementary material for details).

\subsection{Six-patch model}

The Landau  functional for the six-patch model to order $\Delta^4_{E_{1,2}}$ has the form
\begin{equation}
\label{eq:F6p}
\begin{gathered}
\mathcal{F}_{6p} = \alpha_1 \left(|\Delta_{E_1}|^2 + |\Delta_{E_2}|^2 \right) + \beta_1 \left(|\Delta_{E_1}|^2 + |\Delta_{E_2}|^2 \right)^2 + \beta_2 \lvert \Delta_{E_1}^2 + \Delta_{E_2}^2 \rvert^2.
\end{gathered}
\end{equation}
As usual, near a superconducting instability, $\alpha_1 \propto (T - T_c)$ and $\beta_1 >0$. The coupling $\beta_2$ can be of any sign, as long as $\beta_1 + \beta_2 >0$.
Minimizing with respect to amplitudes and phases of $\Delta_{E_1}$ and $\Delta_{E_2}$ we find that for $\beta_2 >0$,  $\mathcal{F}_{6p}$ is minimized by $(\Delta_{E1},\Delta_{E2})=\Delta_E e^{i \phi} (1,\pm i)$ (Refs. \cite{Nandkishore2012,PhysRevB.86.020507}).
  This state breaks $U(1)$ phase symmetry and additionally breaks $Z_2$  time-reversal symmetry. For $\beta_2 <0$, $\mathcal{F}_{6p}$ is minimized by $(\Delta_{E1},\Delta_{E2})=\Delta_E(\cos\gamma,\sin\gamma)$, where $\gamma$ is arbitrary.
 To fix $\gamma$, one needs to include terms of  sixth order in $\Delta_{E_{1,2}}$.
  The relevant sixth-order term is \cite{RevModPhys.63.239,Nandkishore2012,Lin2018,2019arXiv190501702C,Schmalian2018,Venderbos2018}
\begin{equation}
\mathcal{F}^{(6)}_{6p} = \frac{\lambda}{2}\left[(\Delta_{E_1}-i \Delta_{E_2})^3(\bar{\Delta}_{E_1}-i \bar{\Delta}_{E_2})^3 + \text{c.c}\right].
\label{eq:6order}
\end{equation}

 For our six-patch model for electron-doped TBG, we derived $\beta_2$ from the underlying microscopic model and found  $\beta_2>0$, i.e., the SC state is a nodeless chiral superconductor.
   Such a state, dubbed $d \pm id$, has been found in several earlier studies of superconductivity in TBG \cite{You2019npj,Lin2018,Lin2019,2019arXiv190301701C,PhysRevLett.121.087001,PhysRevB.98.085436,PhysRevLett.121.217001,PhysRevB.98.241407,PhysRevB.99.195120}.
It breaks time-reversal symmetry, but does not break $C_3$ lattice rotational symmetry.

  It was argued that nematic fluctuations ~\cite{Kozii2019,Venderbos2018}  in the normal state (more accurately, nematic components of charge or spin density wave fluctuations) do affect $\beta_2$, and if these fluctuations are strong, they can, in principle,  reverse the sign of $\beta_2$ and convert the SC state in the six-patch model into a nematic SC.  We did not analyze the strength of nematic fluctuations in our six-patch model.  Instead we show how a nematic SC state can still develop in the twelve-patch model for hole doping, even if $\beta_2>0$, due to the presence of another superconducting component.

\subsection{Twelve-patch model}

As we demonstrated in Sec.~\ref{sec:gap}, there are two attractive pairing channels for the twelve-patch model -- one-component $A_2$ and two-component $E$ channels.
  Up to fourth order in the gap function, the Landau free energy is
\begin{align}
\mathcal{F}_{12p} &= \alpha_1 \left(|\Delta_{E_1}|^2 + |\Delta_{E_2}|^2 \right) + \alpha_2 |\Delta_{A_2}|^2 + \beta_1 \left(|\Delta_{E_1}|^2 + |\Delta_{E_2}|^2 \right)^2 + \beta_2 \lvert \Delta_{E_1}^2 + \Delta_{E_2}^2 \rvert^2 + \beta_3 |\Delta_{A_2}|^4 \notag\\
&+ \gamma_1 \left(|\Delta_{E_1}|^2 + |\Delta_{E_2}|^2 \right) |\Delta_{A_2}|^2  + \gamma_2 \left[ \left(  \Delta_{E_1}^2 + \Delta_{E_2}^2 \right) \bar{\Delta}_{A_2}^2 + \left(  \bar{\Delta}_{E_1}^2 + \bar{\Delta}_{E_2}^2 \right) \Delta_{A_2}^2 \right]  \notag\\
&+ \delta \left[ \left( 2 \Delta_{E_1} |\Delta_{E_2}|^2 + \bar{\Delta}_{E_1} \Delta_{E_2}^2  -\Delta_{E_1} |\Delta_{E_1}|^2 +  2 \Delta_{E_2} |\Delta_{E_1}|^2 + \bar{\Delta}_{E_2} \Delta_{E_1}^2  -\Delta_{E_2} |\Delta_{E_2}|^2  \right) \bar{\Delta}_{A_2} + \text{c.c.} \right].
\label{FGLfull}
\end{align}
where bar on top of $\Delta$ means complex conjugation, and  $\alpha_1\propto T-T_c^{E}, \alpha_2 \propto T-T_c^{A_2}$ change sign at the  critical temperatures for the pairing in $E$ and $A_2$ channels.
 We find that all prefactors for the fourth-order terms  -- $\beta_1, \beta_2, \beta_3 , \gamma_1, \gamma_2$ and $\delta$,  are positive.

Immediately below the largest of $T_c^E$ and $T_c^{A_2}$,  the system develops either $E$ or $A_2$ superconducting order.
When  $T_c^E$ is larger, we have for the order in the $E$ channel
\begin{equation}
\mathcal{F}_{12p}^E = \alpha_1 \left(|\Delta_{E_1}|^2 + |\Delta_{E_2}|^2 \right) + \beta_1 \left(|\Delta_{E_1}|^2 + |\Delta_{E_2}|^2 \right)^2 + \beta_2 \lvert \Delta_{E_1}^2 + \Delta_{E_2}^2 \rvert^2.
\label{GL2delta}
\end{equation}
 This $\mathcal{F}_{12p}^E$ has the same form as $\mathcal{F}_{6p}$ in the six-patch model.  Like there, we found $\beta_2 >0$.  Then the state immediately below   $T_c^E$ is a nodeless SC, which breaks time-reversal symmetry, but does not break $C_3$ lattice rotational symmetry.

 When $T_c^{A_2}$ is larger, we have for the order in the $A_2$ channel
\begin{equation}
\mathcal{F}^{A_2}_{12p} =  \alpha_2 \Delta_{A_2}^2   + \beta_3 \Delta_{A_2}^4.
\end{equation}
 The $A_2$ order is odd under $C_2$ rotations, but it
 does not break $C_3$ lattice rotation symmetry.

We now consider coexistence states, in which both $A_2$ and $E$ order parameters are non-zero. We see from Eq.~(\ref{FGLfull}) that there are two types of terms in $\mathcal{F}_{12p}$, which contain products of $\Delta_{A_2}$ and $\Delta_{E_{1,2}}$. The terms with
 coefficients $\gamma_1$ and $\gamma_2$ are "conventional" bi-quadratic terms, which in a generic case set relative magnitudes and phases of $A_2$ and $E$ gap components.
  However,  there is the additional term in  Eq.~(\ref{FGLfull}) with prefactor $\delta$, which is  linear in $\Delta_{A_2}$ and cubic in $\Delta_E$.  Such a term  is allowed   by all symmetries.
   Indeed, one can explicitly verify that it is symmetric with respect to an overall $U(1)$ phase rotation and does not change under $C_3$ and $C_2$ rotations.
  For  the invariance under $C_2$, it is essential that
  our choice of
  eigenfunctions ${\vec \Delta}_{E_1}$ and
 ${\vec \Delta}_{E_2}$ transform under $C_2$ as  ${\vec \Delta}_{E_1} \leftrightarrow - {\vec \Delta}_{E_2}$ (see Eq. (\ref{wed_3}) and discussion after it).
The structure of this $\delta$ term is
similar to that of the sixth-order term in Eq.~(\ref{eq:6order}) of the six-patch model.  Indeed, the $\delta$ term can be re-expressed as
\begin{align}
- \frac{\delta}{2} \bar{\Delta}_{A_2} &\Bigg(\left[ (\Delta_{E_1}-i \Delta_{E_2})^2(\bar{\Delta}_{E_1}-i \bar{\Delta}_{E_2})+ (\Delta_{E_1}+i \Delta_{E_2})^2(\bar{\Delta}_{E_1}+i \bar{\Delta}_{E_2}) \right]  \notag \\
&- i\left[ (\Delta_{E_1}-i \Delta_{E_2})^2(\bar{\Delta}_{E_1}-i \bar{\Delta}_{E_2})- (\Delta_{E_1}+i \Delta_{E_2})^2(\bar{\Delta}_{E_1}+i \bar{\Delta}_{E_2}) \right]\Bigg) + \text{c.c.}
\end{align}
 We will show that the $\delta$ term in the twelve-patch model and the sixth-order term in the six-patch model will play a similar role regarding the breaking of lattice rotation symmetry.
 We note in passing that the term cubic in one SC order parameter and linear in the other was recently proposed in Ref.~\cite{Zinkl2019} in the context of chiral $p$- and $f$-wave pairing states on the square lattice, with application to Sr$_2$RuO$_4$.

 To understand the role played by the $\delta$ term, it is  instructive to first consider the structure of the coexistence state without this term, and then add it.  This is what we do next.

\subsubsection{The structure of the coexistence state for $\delta=0$}

Without loss of generality, we choose the phase of complex $\Delta_{A_2}$ to be zero, i.e., set  $\Delta_{A_2}$ to be real.  We parametrize complex $\Delta_{E_1}$ and $\Delta_{E_2}$ as
\begin{equation}
\begin{pmatrix}
\Delta_{E_1} \\
\Delta_{E_2}
\end{pmatrix}=
\Delta_E e^{i\phi_+}
\begin{pmatrix}
e^{i \phi_-} \cos{\gamma}  \\
e^{-i \phi_-} \sin{\gamma}
\end{pmatrix},
\label{gap_param}
\end{equation}
where $\gamma\in[0,\pi/2]$.  Using this parameterization,
we rewrite the Landau free energy Eq.~(\ref{FGLfull})  with $\delta=0$ as
\begin{align}
\mathcal{F}_{12p}^{\delta=0} &=2 \alpha_1 \Delta_{E}^2 + \alpha_2 \Delta_{A_2}^2 +\beta_1 \Delta_E^4 + \beta_3 \Delta_{A_2}^4 + \gamma_1 \Delta_E^2 \Delta_{A_2}^2 + \tilde{\beta}_2  \left( \cos ^2 {2\gamma} + \sin ^2 {2\gamma} \cos^2 2\phi_{-} \right) \notag\\
&+ 2\tilde{\gamma}_2 \left(  \cos {2\phi_{+}} \cos{2\phi_{-}} - \sin {2\phi_{+}} \sin {2\phi_{-}} \cos{2\gamma} \right),
\label{FGLd0}
\end{align}
where $\tilde{\beta}_2= \beta_2 \Delta_E^4, \tilde{\gamma}_2 =  \gamma_2 \Delta_E^2 \Delta_{A_2}^2.$
 Minimizing the functional, we find two types of solutions, one for   $\tilde\gamma_2<\tilde\beta_2$, another for  $\tilde\gamma_2 > \tilde\beta_2$. The first solution is realized when the coexistence state emerges out of the $E$ state, the second is when it emerges from the $A_2$ state.

 For $\tilde\gamma_2 < \tilde\beta_2$ we obtain from minimization
\begin{align}
\cos 2 \phi_{-} &= - \frac{\tilde{\gamma}_2}{\tilde{\beta}_2} \cos 2 \phi_{+} \notag \\
\cos 2 \gamma &= \frac{\tilde{\gamma}_2 \sin 2\phi_{+}}{\tilde{\beta}_2  \sin 2\phi_{-}}.
\label{manifold}
\end{align}
 At $\Delta_{A_2} =0$, ${\tilde \gamma}_2 =0$, and Eq. (\ref{manifold}) yields $\gamma = \pi/4$ and $\phi_- = \pm \pi/4$, as expected for the pure $E$ state.

 Substituting $\cos 2 \phi_{-}$ and $\cos 2 \gamma$ from Eq.~(\ref{manifold})
  into the Landau free energy, we find that
\begin{align}
\mathcal{F}_{12p}^{\delta=0} &=2 \alpha_1 \Delta_{E}^2 + \alpha_2 \Delta_{A_2}^2 +\beta_1 \Delta_E^4 + \beta_3 \Delta_{A_2}^4 + \gamma_1 \Delta_E^2 \Delta_{A_2}^2  - \frac{\tilde{\gamma}_2^2}{\tilde{\beta}_2} \notag \\
&= 2 \alpha_1 \Delta_{E}^2 + \alpha_2 \Delta_{A_2}^2+ \beta_1 \Delta_E^4 + \beta_3 \Delta_{A_2}^4 + \gamma_1 \Delta_E^2 \Delta_{A_2}^2  - \frac{\gamma_2^2 \Delta_{A_2}^4}{\beta_2},
\label{manifold_min}
\end{align}
does not depend on $\phi_+$.  This implies that the order parameter manifold contains, in addition to  $U(1)$ total phase symmetry, another, extra  $U(1)$, associated with the freedom to rotate the common phase of $\Delta_{E_1}$ and $\Delta_{E_2}$ with respect to $\Delta_{A_2}$.  In addition,  Eq. (\ref{manifold}) for fixed $\phi_+$ allows two solutions ($\phi_-, \gamma)$ and $(-\phi_-, \pi/2 - \gamma)$.  One solution transforms into the other if we interchange $\Delta_{E_1}$ into $\Delta_{E_2}$.  The  full order parameter manifold is then $U(1) \times U(1) \times Z_2$.
One can verify that this $Z_2$ is associated with time-reversal symmetry.

\begin{figure}[tb]
\center{\includegraphics[width=0.65\linewidth]{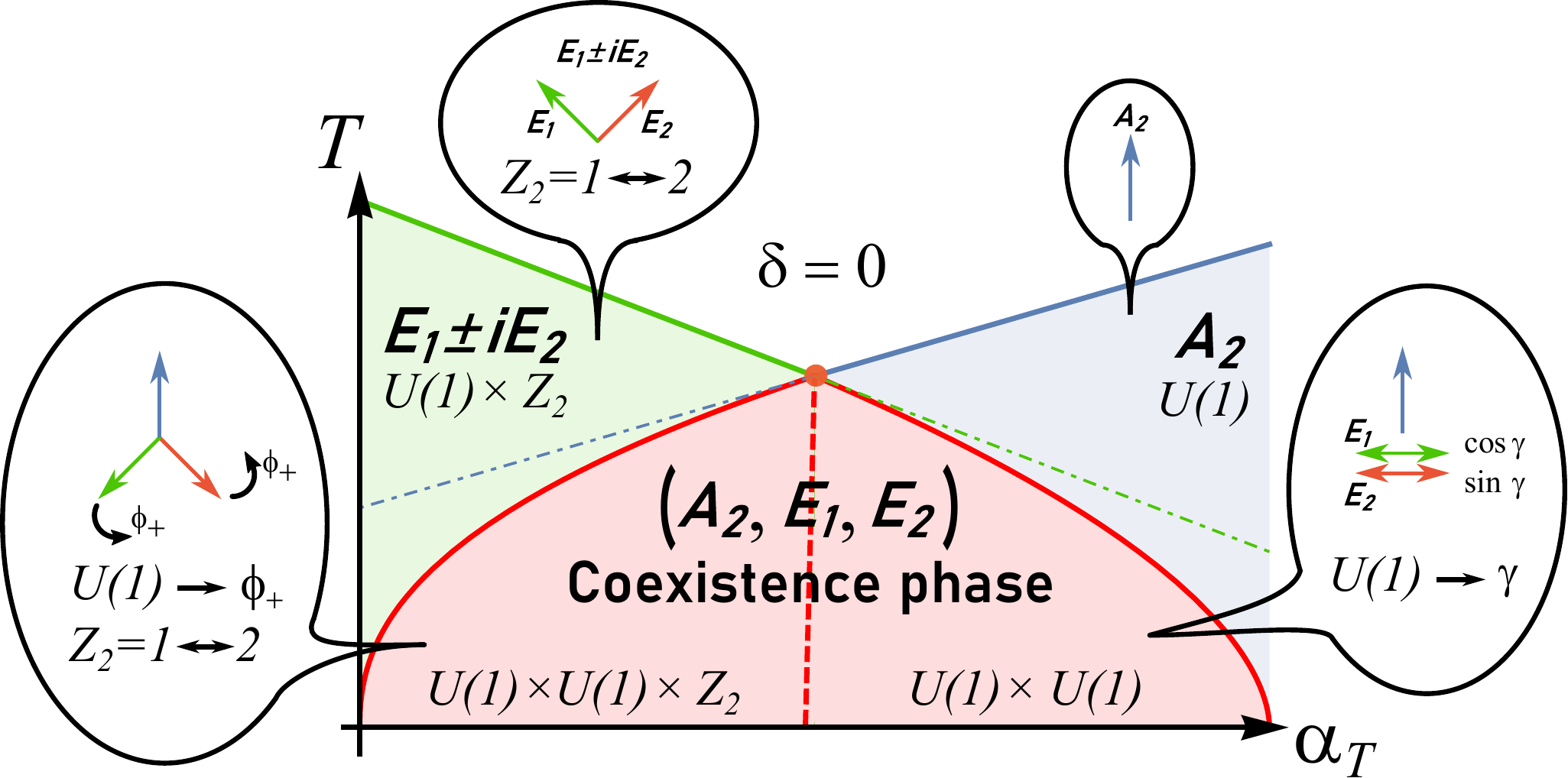} }
\centering{}\caption{A schematic phase diagram for
 the special case $\delta= 0$ in Eq. (\ref{FGLfull}) for the free energy.
  We use the strength of the non-local interactions $\alpha_T$ as tuning parameter. The orange circle marks the point where $A_2$ and $E$ channels are exactly degenerate.
    In the green shaded region the superconducting state is  pure $E$  ($E_1 \pm i E_2$), in the blue shaded region the SC
     is pure  $A_2$. In the red shaded region,  both $E=(E_1, E_2)$ and  $ A_2$ gap components are non-zero. The  dashed red line marks a phase transition between two coexistence states with different order parameter manifolds.   On the left of this line the manifold is $U(1)\times U(1)\times Z_2$, on the right it is
        $U(1)\times U(1)$. In both cases, the order parameter manifold contains an additional continuous $U(1)$  symmetry.   The insets illustrate symmetry operations.  The directions of blue, green, and brown arrows  correspond to the phases of $A_2,E_1,$ and $E_2$  order parameters.}
\label{phdiagd0}
\end{figure}

For $\tilde\gamma_2 > \tilde\beta_2$,  the solution Eq. (\ref{manifold})
disappears. The new minima are at
\begin{align}
\begin{cases}
\phi_+&=0 \\
 \phi_-&=\pm\pi/2
 \end{cases}
 \; \text{and} \;
 \begin{cases}
 \phi_+&=\pm\pi/2 \\
  \phi_-&=0
  \end{cases}
 \label{eq:R0}
\end{align}
Substituting these solutions into the  Landau free energy, we obtain that it does not depend on $\gamma$:
\begin{align}
\mathcal{F}_{12p}^{\delta=0} &=2 \alpha_1 \Delta_{E}^2 + \alpha_2 \Delta_{A_2}^2 +\beta_1 \Delta_E^4 + \beta_3 \Delta_{A_2}^4 + \gamma_1 \Delta_E^2 \Delta_{A_2}^2  +\tilde\beta_2-2\tilde\gamma_2 \notag \\
&= 2 \alpha_1 \Delta_{E}^2 + \alpha_2 \Delta_{A_2}^2+ \beta_1 \Delta_E^4 + \beta_3 \Delta_{A_2}^4 + \gamma_1 \Delta_E^2 \Delta_{A_2}^2  +\beta_2\Delta_E^4-2\gamma_2\Delta_E^2\Delta_{A_2}^2
\label{manifold_min_1}
\end{align}
This means that the order parameter manifold again has an additional continuous $U(1)$ symmetry.  To obtain the full order parameter manifold in this case, we note that the  four solutions in Eq.~(\ref{eq:R0}) can be re-expressed as
\begin{equation}
\begin{pmatrix}
\Delta_{E_1} \\
\Delta_{E_2}
\end{pmatrix}=
i \Delta_E
\begin{pmatrix}
 \cos{\gamma}  \\
 \sin{\gamma}
\end{pmatrix},
\label{gap_param_1}
\end{equation}
 if  we allow $\gamma$ to vary between zero and $2\pi$. This implies that the order parameter manifold for $\tilde\gamma_2 > \tilde\beta_2$  is $U(1) \times U(1)$. There is no additional $Z_2$, because the phase of $\Delta_{E_1}$ and $ \Delta_{E_2}$ in (\ref{gap_param_1}) is either the same or differs by $\pi$, in which case  phase reversal does not create a distinct SC state.  Put differently, the interchange $\Delta_{E_1} \leftrightarrow \Delta_{E_2}$ can be absorbed into a  variation of $\gamma$.

We show the phase diagram for $\delta=0$ and sketches of the gap configurations  in Fig. \ref{phdiagd0} with $\alpha_T$ as a tuning parameter.
Along the transition line at $\tilde\gamma_2 = \tilde\beta_2$, one of the $E$ components vanishes, and the order parameter manifold reduces to $U(1) \times Z_2$.

The existence of the continuous $U(1)$ symmetry in the order parameter manifold is highly unusual.
In general, one would expect only one $U(1)$ to be present, associated with the symmetry with respect to rotations  of the common phase. We will see below that  in the presence of the $\delta$ term, the continuous $U(1)$ symmetry  is replaced by a discrete $C_3$ symmetry.

\subsubsection{The structure of the coexistence state for nonzero $\delta$}

Next, we consider the full Landau free energy, Eq.~(\ref{FGLfull}), with the $\delta$-term.
 We use the same parametrization as in Eq.~(\ref{gap_param}).
The full analysis of Eq. (\ref{FGLfull}) is rather cumbersome, but the outcome can be understood by just expanding near the boundaries of the coexistence phase.
 Near the left boundary, where  $\Delta_{A_2}\ll\Delta_E$, we have at $\delta =0$ $\gamma = \pi/4$ and  $\phi_- = \pm \pi/4$.
 Accordingly, at finite $\delta$, we set $\gamma=\pi/4+\epsilon_\gamma$
 and $\phi_-= \pm (\pi/4+\epsilon_-)$,
  where $\epsilon_\gamma, \epsilon_- \sim  \delta \Delta_{A_2}/\Delta_E \ll 1$.
   We will see below that this expansion is  valid for $\delta^2 < 2 \beta_2 \gamma_2$.
   The solutions with opposite sign of $\phi_-$ transform into each other under $\Delta_{E_1} \leftrightarrow \Delta_{E_2}$, i.e., the order parameter manifold contains
    $Z_2$ associated with time reversal, like for $\delta =0$.

  Substituting this expansion into (\ref{FGLfull}) and minimizing with respect to $\epsilon_\gamma,\epsilon_-$, we obtain to leading order in $\Delta_{A_2}/\Delta_E$:
\begin{align}
\epsilon_\gamma&= \frac{\delta}{2\beta_2}\frac{\Delta_{A_2}}{\Delta_E} \sin{\phi_+} \notag \\
\epsilon_-&=\frac{\delta}{2\beta_2}\frac{\Delta_{A_2}}{\Delta_E} \cos{\phi_+}
\label{left_sol}
\end{align}
 Substituting these expressions back into the Landau free energy we obtain
 \begin{align}
\mathcal F_{12p} =\mathcal F_{12p}^{\delta=0} - \frac{\delta^2}{\beta_2}\Delta_E^2\Delta_{A_2}^2
- \delta\frac{2\beta_2\gamma_2-\delta^2}{\beta_2^2}\cos3\phi_+ \Delta_E \Delta_{A_2}^3 + \delta^2\frac{8\beta_2\gamma_2-3\delta^2}{4\beta_2^3}\Delta_{A_2}^4.
\label{eq:FL0}
\end{align}
 We see that the free energy now depends on $\phi_+$ via the $\cos3\phi_+$ term.  For $\delta^2 < 2 \beta_2 \gamma_2$, minimization with respect to $\phi_+$ yields
three solutions  $\phi_+ = (0, 2\pi/3, - 2 \pi/3)$.
We  see  that the $\delta$ term  reduces the additional continuous $U(1)$ symmetry to a discrete $C_3$ symmetry. The system spontaneously chooses one out of three allowed values of $\phi_+$, and thereby breaks lattice rotational symmetry and becomes a nematic superconductor.  Note that one of the states has $\phi_+=0$ and, hence, $\gamma = \pi/4$.  For this state, the magnitudes of $\Delta_{E_1}$ and $\Delta_{E_2}$ are equal, only the relative angle $2\phi_-$ varies with $\Delta_{A_2}$.
However, the two $E$ components of the gap are not equal in any given patch,
 as one gets multiplied by $\cos (4 \theta_i + 3\pi/4)$, and the other by  $\sin(4 \theta_i + 3\pi/4)$, where, we remind, $\theta_i$ specify the directions towards VH points.
  For the other two solutions ($\phi_+=\pm2\pi/3$), we verified that the $E$ components of the gap are the same as for the first solution if we rotate $\theta_i$ by $\pm
 \pi/3$.
 For $\delta^2 > 2 \beta_2 \gamma_2$,  the $\phi_+$-dependent term in the free energy Eq.~\eqref{eq:FL0} changes sign. In this case, another solution, with
  $\phi_-$  approximately $\pm3\pi/4$ for $\Delta_{A_2}\ll\Delta_E$, becomes energetically favorable.

We also note that (i) the prefactor for the term quadratic in $\Delta_{A_2}$ in Eq.~(\ref{eq:FL0}) is negative, i.e., for non-zero $\delta$ the transition temperature into the coexistence state is larger than the original $T_c^{A_2}$, where $\alpha_2$ in Eq.~(\ref{FGLfull}) changes sign and (ii) the free energy (\ref{eq:FL0}) has a term proportional to $\Delta^3_{A_2}$. This term renders the transition between the pure $E_2$ state and the coexistence state first order.

 We consider next the situation near the right boundary of the coexistence phase, where $\Delta_E \ll \Delta_{A_2}$.  Let us assume for definiteness that without the $\delta$-term,  $\phi_-=\pi/2$ and $\phi_+ =0$  ($\Delta_{E_1} = i \cos{\gamma}, \Delta_{E_-} =-i \sin{\gamma}$), cf. Eq.~\eqref{eq:R0}.  When $\delta$ is non-zero,  we expand
 $\phi_- = \pi/2 + \varepsilon_-$ and $\phi_+=0+\varepsilon_+$.
 Minimizing with respect to $ \varepsilon_{\pm}$, we obtain
\begin{align}
\varepsilon_+ &= \frac{\delta}{8\gamma_2}\frac{\Delta_E}{\Delta_{A_2}}\frac{\sin\gamma-\cos\gamma}{\sin\gamma\cos\gamma} \notag\\
\varepsilon_- &=\frac{\delta}{4\gamma_2}\frac{\Delta_E}{\Delta_{A_2}}(\cos3\gamma-\sin3\gamma),
\label{right_sol}
\end{align}
and at the minimum
 \begin{equation}
 \mathcal{F}_{12p} = 2 \alpha_1 \Delta_{E}^2 + \alpha_2 \Delta_{A_2}^2 +  \left(\beta_1 + \beta_2 - \frac{\delta^2}{2 \gamma_2} \right) \Delta_E^4 + \beta_3 \Delta_{A_2}^4 + (\gamma_1 - 2 \gamma_2) \Delta_E^2 \Delta_{A_2}^2 + \frac{\beta_2 \delta^2 \left(\sin (6\gamma) -1 \right) \Delta_E^6}{4 \gamma_2^2 \Delta_{A_2}^2}.
 \label{eq:F0R}
 \end{equation}
 Contrary to the  previous case, there is no $U(1)$ breaking term at order $O(\delta)$.
 However, such a term appears at order  $\delta^2$ with the structure $\delta^2 \sin{6 \gamma}$.
  Minimizing with respect to $\gamma$, we obtain $\gamma = \pi/4 + \pi n/3$, where $n = 0,\ldots,5$ is an integer.
We observe that now we have six solutions for $\gamma$ within a $2\pi$ interval.
  One can verify that out of these six solutions, three are time-reversal partners of the other three, i.e., time-reversal symmetry is broken.
   One can understand this on physical grounds, because once  the phase difference $2 \phi_-$ between
  $\Delta_{E_1}$ and $\Delta_{E_2}$ becomes different from $\pi$, $\phi_-$ and $- \phi_-$  describe non-identical gap configurations, hence under  time-reversal the system transforms into a physically different state.
     The remaining three solutions transform into each other under elements of $C_3$,
      i.e., the order parameter manifold is $U(1) \times C_3 \times Z_2$, the
     same that we obtained near the left boundary of the coexistence phase.
    Note that in Eq.~\eqref{eq:F0R} the correction to $\alpha_1$ vanishes, and there is no $\Delta^3_E$ term.  As a consequence, the transition from the pure $A_2$ state into the coexistence state is second order as long as
 $4\left(\beta_1 + \beta_2 - \frac{ \delta^2}{2\gamma_2} \right) \beta_3  > ( \gamma_1 -2 \gamma_2)^2$
(see Refs. \cite{Liu1973,PhysRevB.67.054505,PhysRevE.88.042141}).

We verified that near the left boundary of the coexistence state,  $\phi_-$ increases with $\Delta_{A_2}$ (cf. Eq.~\eqref{left_sol}),
and near the right boundary $\phi_-$ decreases as $\Delta_E$ increases (cf. Eq.~\eqref{right_sol}),
  i.e.,  $\Delta_{E_1}$ and $\Delta_{E_2}$ rotate towards  each other.
     This strongly suggests that the gap structure in the coexistence state evolves continuously for small, but non-zero $\delta$.
  We solved numerically for the gap at arbitrary ratio of $\Delta_{A_2}/\Delta_E$ and found that this is indeed the case if $\delta^2 < 2 \beta_2 \gamma_2$.
   Specifically,  for the "symmetric" state with  $\phi_+=0$ and $\gamma=\pi/4$, we found
a continuous change of  $\phi_-$ inside the coexistence phase from
$\phi_- \sim \pi/4$
for $\Delta_{A_2}\ll\Delta_E$ to $\phi_- \simeq \pi/2$ for $\Delta_{A_2}\gg\Delta_{E}$.
   We show the phase diagram in Fig.~\ref{phdiag1} along with the structure of the pure and coexistence states.

\subsubsection{The case of large $\delta$}

 We now show that a new state emerges at $\delta^2 > 2 \beta_2 \gamma_2$, which breaks $C_3$ symmetry, but preserves time-reversal symmetry.  To see this, we look again at the  solutions close to the left and right boundaries.
 We found  before that one of the solutions from the $C_3$ manifold is a symmetric one: $\phi_+=0$ and $\gamma = \pi/4$, i.e., $|\Delta_{E_1}| = |\Delta_{E_2}|$.  Let us keep these values of $\phi_+$ and $\gamma$, but not assume that $\Delta_{A_2}/\Delta_E$ is small and treat $\phi_-$ as parameter.
 We will use this as an ansatz for the ground state for larger $\delta$ and then verify that it is a stable minimum.

 Substituting into  Eq. (\ref{FGLfull}), we obtain
\begin{align}
\mathcal{F}_{12p} &= 2 \alpha_1 \Delta_E^2 + \alpha_2 \Delta_{A_2}^2 + 4 \beta_1 \Delta_E^4  +  \beta_3 \Delta_{A_2}^4 + 2 \gamma_1 \Delta_E^2 \Delta_{A_2}^2 \notag\\
&  + \cos(2\phi_-) \left(2 \gamma_2 \Delta_E^2 \Delta_{A_2}^2 + \beta_2 \Delta_E^4  + 2 \sqrt{2} \delta \Delta_E^3 \Delta_{A_2} \cos(\phi_-) \right).
\label{GLsimple}
\end{align}

One can check that at large enough $\delta$, the free energy has smallest value when $\phi_- = \pm \pi$.  For this $\phi_-$, Eq. (\ref{GLsimple}) reduces to
\beq
\mathcal{F} = 2 \alpha_1 \Delta_E^2 + \alpha_2 \Delta_{A_2}^2 + \left( 4 \beta_1 + \beta_2 \right) \Delta_E^4  +   \beta_3 \Delta_{A_2}^4
  + 2 (\gamma_1 +\gamma_2)\Delta_E^2 \Delta_{A_2}^2   - 2\sqrt{2} \delta \Delta_E^3 \Delta_{A_2}.
 \label{th_1}
 \eeq
 In such a state the phase of the two $E$ components of the gap is opposite to the phase of the $A_2$ component, i.e., all three gap components, viewed as vectors, are directed along the same axis.
 Such a state preserves $Z_2$ time-reversal symmetry.

 Rotational $C_3$ symmetry requires that there must be two other states with the same energy.
In total, we find
\begin{align}
\phi_+&=0 \quad \gamma=\pi/4 \quad \phi_-=\pm \pi\notag\\
\phi_+&=\pi/2 \quad \gamma=5\pi/12 \quad \phi_-=\pi/2 \notag\\
\phi_+&=\pi/2 \quad \gamma=\pi/12 \quad \phi_-=-\pi/2 .
\end{align}

We now analyze where this "collinear" state is located in the phase diagram. For this we assume that it is present for some $\Delta_{A_2}$ and $\Delta_E$ and check its stability. For definiteness we choose the "symmetric" state with $\phi_+ =0, \phi_- = \pi$, $\gamma = \pi/4$ and vary the angles by $\phi_- = \pi + \varepsilon_-$, $\phi_+=\varepsilon_+$ and $\gamma=\pi/4+\varepsilon_\gamma$.
    Substituting this into the free energy, we obtain to second order in $\varepsilon_i$
\begin{align}
\mathcal{F}_{12p} &= 2 \alpha_1 \Delta_E^2 + \alpha_2 \Delta_{A_2}^2 + \left( 4 \beta_1 + \beta_2 \right) \Delta_E^4  +   \beta_3 \Delta_{A_2}^4 \notag\\
&  + 2 (\gamma_1 +\gamma_2)\Delta_E^2 \Delta_{A_2}^2   - 2\sqrt{2} \delta \Delta_E^3 \Delta_{A_2} \notag\\
&+ \left[5\sqrt{2} \delta \Delta_E^3 \Delta_{A_2} - 4 \gamma_2 \Delta_E^2 \Delta_{A_2}^2 - 4 \beta_2 \Delta_E^4 \right] \varepsilon^2_- +9\sqrt{2}\delta\Delta_{A_2}\Delta_E^3\varepsilon^2_\gamma +\left[ \sqrt{2}\delta\Delta_E^3\Delta_{A_2}-4\gamma_2\Delta_E^2\Delta_{A_2}^2 \right]\varepsilon^2_+.
\end{align}
 The stability conditions are then
\begin{align}
\sqrt{2}\delta\Delta_E^3\Delta_{A_2}-4\gamma_2\Delta_E^2\Delta_{A_2}^2\geq&0\notag\\
5\sqrt{2} \delta \Delta_E^3 \Delta_{A_2} - 4 \gamma_2 \Delta_E^2 \Delta_{A_2}^2 - 4 \beta_2 \Delta_E^4 \geq& 0,
\end{align}
These conditions set the boundaries of the collinear phase at
\begin{align}
\sqrt{2}\frac{5 \delta + \sqrt{25 \delta^2 - 32 \beta_2 \gamma_2} }{8 \beta_2} \Delta_{A_2}\geq \Delta_E \geq \sqrt{2}\frac{5 \delta - \sqrt{25 \delta^2 - 32 \beta_2 \gamma_2} }{8 \beta_2} \Delta_{A_2}
\end{align}
and
\begin{equation}
\Delta_E \geq \frac{4\gamma_2}{\sqrt{2}\delta}\Delta_{A_2}.
\end{equation}
 The first boundary is where fluctuations near $\phi_- = \pi$ become unstable, the second is where fluctuations near $\phi_+ = 0$ become unstable.
Fluctuations of $\gamma$ do not give an additional constraint.
Combining these two conditions, we obtain that the phase with unbroken time-reversal symmetry exists once $\delta^2$ exceeds
$2 \beta_2 \gamma_2$ (we need $\sqrt{2}[5 \delta + \sqrt{25 \delta^2 - 32 \beta_2 \gamma_2} ] /8 \beta_2 \geq 4\gamma_2/\sqrt{2}\delta$).
It starts as a line in the phase diagram at
$\delta^2 = 2 \beta_2 \gamma_2$ and expands into the coexistence  phase for larger $\delta$.
We show the phase diagram at large $\delta$ in Fig. \ref{phdiag1}
along with the states from the $C_3$ manifold inside the collinear phase, and in Fig. \ref{total_gaps} we present the plots of the total gap function
$\Delta^{SC}_{12p}=\Delta_{A_2}\vec\Delta_{12p}^{A_2} + \Delta_{E_1}\vec\Delta_{12p}^{E_1} + \Delta_{E_2}\vec\Delta_{12p}^{E_2} $ for the three regions within the coexistence phase in the right panel of Fig. \ref{phdiag1}.

The condition $\delta^2\geq2\beta_2\gamma_2$ coincides with the condition that $\phi_-$ near the left boundary of the coexistence phase
  jumps from $\pi/4$ to $3\pi/4$.
It then further increases with  $\Delta_{A_2}$ and reaches $\pi$  at the left boundary of the state with unbroken time-reversal symmetry.
The evolution of the gap between the right boundary of the coexistence state and the collinear phase is
 more involved
 and we refrain from discussing it in detail.
  We note in passing that there is a certain analogy between the phase diagram and excitations in our case and for a 2D Heisenberg antiferromagnet in a magnetic field,
 whose phase diagram also contains an intermediate up-up-down phase with
 collinear ordering of spins in the three sublattices \cite{Chubukov_1991}.

\section{Gap structure along the full Fermi surface and experimental consequences}

\begin{figure}[t]
\includegraphics[width=0.99\linewidth]{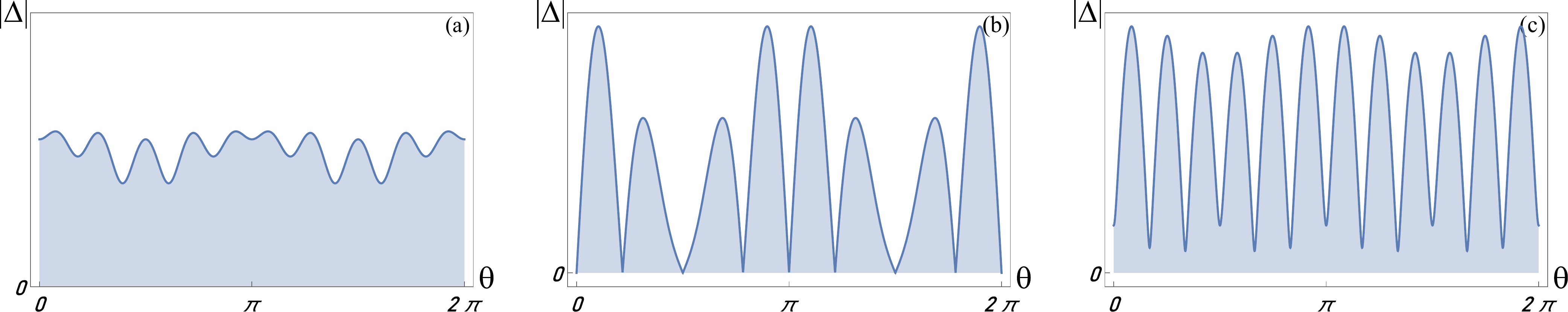} \quad
\centering{}\caption{The magnitude  of the total gap function
$\Delta^{SC}_{12p}=\Delta_{A_2}\vec\Delta_{12p}^{A_2} + \Delta_{E_1}\vec\Delta_{12p}^{E_1} + \Delta_{E_2} \vec\Delta_{12p}^{E_2}$ along the Fermi surface, when $\vec\Delta_{12p}^{A_2}, \vec\Delta_{12p}^{E_1}$, and
$\vec\Delta_{12p}^{E_2}$ are viewed as functions of continuous $\theta$ rather than of $\theta_i$ at Van Hove points.
The three panels correspond to three coexistence states in  the right panel of Fig. \ref{phdiag1}.  We have chosen the symmetric state with $|\Delta_{E_1}| = |\Delta_{E_2}|$ (one of the states in $C_3$ manifold).
 Panel (b) is for the "collinear" state in the middle of the right panel of Fig.  \ref{phdiag1}, and panels (a) and (c) are for the states to the left and to the right of the collinear state, respectively.
We used  $\Delta_{A_2}/\Delta_E = 0.2, \phi_{-} = 0.86$ in panel (a),  $\Delta_{A_2}/\Delta_E = 0.71$ in panel (b), and  $\Delta_{A_2}/\Delta_E = 6.6, \phi_{-} = 1.26$ in panel (c).
  The gap functions in panels (a) and (c) are complex numbers, and $|\Delta^{SC}_{12p}|$ has no nodes.
   The gap function in the collinear phase is real and has nodes, because $\vec\Delta_{12p}^{E_1}$, $\vec\Delta_{12p}^{E_2}$ and $\vec\Delta_{12p}^{A_2}$ have nodes (see Fig. \ref{3gapsol}).    The gap structure in the left panel of Fig. \ref{phdiag1} is the same as in panels (a) and (c).
   }
\label{total_gaps}
\end{figure}

 The gap structure along the full Fermi surface
  is    shown in the three panels (a)-(c) in Fig. \ref{total_gaps}  for the
 three regions of the phase diagram in the right panel of Fig. \ref{phdiag1}
   (the gap structure for the phase diagram in  the left panel of Fig. \ref{phdiag1} is the same, but without middle panel (b)).
   The gap function in  panel (b) is for a SC state which preserves time-reversal symmetry, and has nodes.
The variation of this gap function  with the angle $\theta$ along the Fermi surface is
\begin{equation}
\Delta (\theta) = \Delta_{A_2} \left( \sin 6\theta + \frac{\Delta_{E}}{\Delta_{A_2}} \sqrt{2} \sin 4\theta \right).
\end{equation}
The number of nodes depends on the ratio $\Delta_{E}/\Delta_{A_2}$: for $\Delta_{E}/\Delta_{A_2}\lesssim 3/(2\sqrt{2})$
there are twelve nodes, for $\Delta_{E}/\Delta_{A_2}\gtrsim 3/(2\sqrt{2})$
 the number of nodes is reduced to eight. The positions of four nodes are protected by time-reversal symmetry and are fixed at $\theta=0, \pi/2, \pi, 3\pi/2$ (at these points both $\sin 6 \theta$ and $\sin 4 \theta$ components vanish).  The location of the other nodal points depends on the ratio $\Delta_{E}/\Delta_{A_2}$.

 The gap functions in panels  (a) and (c) are for the states with  broken time-reversal symmetry. These gap functions are nodeless by obvious reasons.
 They can be parameterized by
\begin{equation}
\Delta (\theta) = \Delta_{A_2} \left[ \sin 6\theta -\frac{1}{\sqrt{2}} \frac{\Delta_{E}}{\Delta_{A_2}} e^{i\phi_+}\left(\cos\phi_- \left[ \cos(4\theta-\gamma)+\sin(4\theta-\gamma)\right]+i\sin\phi_- \left[\cos(4\theta+\gamma)+\sin(4\theta+\gamma) \right]
\right)\right],
\end{equation}
 where $\phi_+,\phi_-$ and $\gamma$ evolve as functions of the ratio $\Delta_{E}/\Delta_{A_2}$  (see Eqs.~\eqref{left_sol} and \eqref{right_sol}).
 The magnitude of the angle variation of $\Delta (\theta)$ depends on  the ratio   $\Delta_{A_2}/\Delta_E$,
and is larger in panel (c) [i.e., for the state at a larger $\alpha_T$ in the right panel of Fig. \ref{phdiag1}].

The gap structures, presented in Fig.  \ref{total_gaps}, can be probed experimentally, by, e.g.,
 QPI analysis of  STM data,
 and ARPES experiments.
  Therefore, they are testable predictions of our theory.  The states with and without nodes can also be distinguished by other techniques, e.g., by measuring the flux penetration depth.   The ratio of $\Delta_{A_2}$ and $\Delta_E$ likely can be varied by, e.g., changing the twist angle or adding uniform strain, which changes the degree of the non-locality of the interactions and hence affects our parameter $\alpha_T$ in Fig.~\ref{phdiag1}.
 A discrete $C_3$ symmetry breaking was reported in Ref.~\cite{kitp_talk} and motivated our study. It can also be detected in STM studies and in  
   angle-resolved photoemission spectroscopy with nanoscale resolution~\cite{nanoARPES}.  A time-reversal symmetry breaking can be detected via a broad range of probes~\cite{Babaev3},
    including measurements of Kerr rotation  \cite{KAPITULNIK2015151}
    and zero-field muon-spin relaxation, which detects weak internal magnetic
fields produced by spontaneous currents, generated around impurities by time-reversal breaking superconducting order~\cite{Mahyari2014,Xiao2012,GARAUD201763,SZLin2016}.
 Domain walls in such superconductors have  magnetic signatures that could be detected
in scanning SQUID and Scanning Hall probe microscope measurements~\cite{Babaev2}. It was also  proposed~\cite{Babaev2017}  that a nematic superconductor possesses topological skyrmions (bound states of two spatially separated half-quantum vortices), which can be detected by STM.

On a qualitative level,  a high $T_c/T_F$ ratio, observed in magic-angle twisted bilayer graphene \cite{Cao2018unconventional},  is more consistent with the existence of attractive pairing interactions  at the bare level rather than with Kohn-Luttinger scenario, in which the attraction develops at second order in the interaction and is likely much weaker.

\section{Conclusions}
\label{sec:concl}

In this paper we performed a comprehensive analysis of superconductivity near Van Hove (VH) filling  in twisted bilayer graphene (TBG) within an itinerant approach. The key motivation for our study has been the recent experimental finding~\cite{kitp_talk,stm2019} that the superconducting order in hole-doped TBG near $n =-2$ breaks $C_3$ lattice rotational symmetry, i.e., the SC state is also nematic.

 We used as an input the effective tight-binding Hamiltonian for the moir\'e superlattice, which describes flat bands \cite{Yuan2018,Koshino2018PRX}. We argued that there are at least two VH fillings, one for hole doping, the other for electron  doping.  At VH filling  for electron doping, there are six VH points, located along high symmetry directions in the Brillouin zone, but away from the zone  boundary. At VH filling for hole doping, there are twelve VH points. They are symmetry related, but each is located away from symmetry directions and the zone boundary.
 We derived effective six-patch and twelve-patch  models for fermions near VH points and projected the interactions into the pairing channel. For the six-patch model, there are two symmetry-allowed pairing interactions
 in the spin-singlet channel.
  For the twelve-patch model this number is five.
  We obtained the values of the interactions by matching the patch models with the microscopic model of Kang and Vafek~\cite{Kang2018strong}, which contains both local (Hubbard) and non-local interactions.
  The relative strength of the non-local interaction is measured by the parameter $\alpha_T$, which was
  estimated to be around 0.23     in TBG.
  We argued that for this $\alpha_T$, the non-local interactions give rise to attraction in certain channels already at the "bare" level, i.e., without including corrections to the pairing interaction from the particle-hole channel.
  In other words, the reconstruction of the band structure due to the twist and projection onto the nearly flat bands leads to attractive pairing interactions in hole-doped TBG.
   The attraction exists for both spin-singlet and spin-triplet channels. We concentrate on the first because experiments on TBG point to spin-singlet pairing~\cite{Cao2018unconventional}.

The symmetry of the superconducting order parameter can be classified based on the irreducible representations of the lattice rotation symmetry group $D_3$. They include two one-dimensional representations, $A_1$ and $A_2$, and one two-dimensional representation, $E$.  Each representation contains an infinite set of different eigenfunctions, but most  become indistinguishable within patch models.
  For the six-patch model, we found that the relevant eigenfunctions are a constant ($s$-wave) from $A_1$ and $d-$wave-like $(\cos{2\theta_i}, \sin{2 \theta_i})$ from $E$, where $\theta_i$ set the directions towards six VH points.
 We found that the interaction in the $E$ channel is attractive and gives rise to $d \pm id$ SC order. It breaks time-reversal symmetry, but preserves $C_3$ lattice rotational symmetry. This agrees with earlier results for the six-patch model~\cite{Lin2019,Isobe2018PRX} and with earlier studies of single-layer graphene around VH filling \cite{Nandkishore2012,PhysRevB.86.020507}.

 For the twelve-patch model, we found four different pairing channels: one in $A_1$, with a constant eigenfunction, one in $A_2$, with an eigenfunction changing signs between neighboring patches, and two in $E$ with eigenfunctions $(\cos{2\theta_i}, \sin{2 \theta_i})$ and $(\cos{4\theta_i}, \sin{4 \theta_i})$.   We found that $A_{2}$ and the $4 \theta$ $E$ channel are attractive, and that for
 realistic $\alpha_T$ the coupling constants in the two channels have near-equal magnitudes.
   We showed that pure $A_2$ order breaks only $U(1)$ phase symmetry, and pure $E$ order is similar to that in the six-patch model, i.e., it breaks $U(1)$ and $Z_2$ time-reversal symmetry, but preserves $C_3$.

 Our key result is that in the coexistence state, where both
  $E$
   and $A_2$ order parameters
   are non-zero, $C_3$ symmetry is broken.
   We argued that this happens due to two reasons: (i) conventional biquadratic couplings between $E$ and $A_2$  order parameters  do not specify the coexistence state, and the order parameter manifold
   has an extra
    $U(1)$ symmetry,   in addition to the  $U(1)$ total phase symmetry, and (ii) the Landau free energy to quartic order contains
   a symmetry allowed term, which is linear in $\Delta_{A_2}$ and qubic in $\Delta_{E_{1,2}}$.
     This term breaks the
      extra  $U(1)$ symmetry  down to threefold  $C_3$.  The system spontaneously chooses one of three equivalent states from $C_3$ manifold and by doing this breaks $C_3$.
As a result, the  coexistence state turns out to be a nematic superconductor.
 We found two phases with broken $C_3$. In one, time-reversal symmetry is also spontaneously broken.  In the other, it is preserved.

Our results present a scenario  for the breaking of
threefold lattice rotation symmetry  in the superconducting state of hole-doped TBG  near $n =-2$, where nematic superconductivity  has been observed~\cite{kitp_talk,stm2019}. We also consider it as a generic, symmetry-based mechanism how a superconductor can break lattice rotational symmetry.
We also emphasize that although in our scenario the nematic long-range order emerges only in the coexistence superconducting phase, nematic order generally survives in some range outside the
 coexistence phase, and nematic fluctuations are strong in the whole region where the pairing susceptibility is enhanced in both $E$ and $A_2$ channels.

\section{Acknowledgments}

We thank E. Andrei, M. Christensen, R. Fernandes, L. Fu, D. Goldhaber-Gordon, P. Jarillo-Herrero, J. Kang,  A. Klein, L. Levitov, M. Navarro Gastiasoro, J. Schmalian, D. Shaffer, O. Vafek, J. Venderbos, and A. Vishwanath for fruitful discussions.
 The work was supported by  U.S. Department of Energy, Office of Science, Basic
Energy Sciences, under Award No. DE-SC0014402.
A.V.C. is thankful to Aspen Center for Physics  (ACP) for hospitality during the completion of this work.
ACP is supported by NSF grant PHY-1607611.

\bibliography{biblio}

\clearpage

\begin{center}
\textbf{\large Supplemental Material}
\end{center}
\setcounter{section}{0}
\setcounter{equation}{0}
\setcounter{figure}{0}
\setcounter{table}{0}
\setcounter{page}{1}
\makeatletter
\renewcommand{\thesection}{S\arabic{section}}
\renewcommand{\theequation}{S\arabic{equation}}
\renewcommand{\thefigure}{S\arabic{figure}}

\section{Diagonalization of single-particle Hamiltonian and the introduction of patch operators}
\label{app:bands}

We introduce the following lattice vectors (see Fig. \ref{lattice})
\begin{eqnarray}
\label{eq_vec1}
\vec{a}_1 = \frac{1}{2} (-1, \sqrt{3}), \;\vec{a}_2 = \frac{1}{2} (-1, -\sqrt{3}), \; \vec{a}_3 = (1, 0), \\
\label{eq_vec2}
\vec{b}_1 = \frac{3}{2} (-1, \sqrt{3}), \;\vec{b}_2 = \frac{3}{2} (-1, -\sqrt{3}), \; \vec{b}_3 = (3, 0), \\
\label{eq_vec3}
\vec{L_1} = \frac{1}{2} (3, \sqrt{3}), \; \vec{L}_2 = (0, \sqrt{3}).
\end{eqnarray}
and Fourier-transform the Hamiltonian Eq.~\eqref{2H}. As a result, we obtain
\begin{equation}
H_{TB}=\sum_{\vec{k}}
\begin{pmatrix}
c_{xA}^{\dag}
c_{xB}^{\dag}
c_{yA}^{\dag}
c_{yB}^{\dag}
\end{pmatrix}
\begin{pmatrix}
T_d & T_{sd1}^* & i T_{sd2} & 0 \\
T_{sd1} & T_d & 0 & i T_{sd2} \\
-i T_{sd2} & 0 & T_d & T_{sd1}^* \\
0 & -i T_{sd2} & T_{sd1} & T_d
\end{pmatrix}
\begin{pmatrix}
c_{xA} \\
c_{xB} \\
c_{yA} \\
c_{yB} \,
\end{pmatrix},
\label{TBHstand}
\end{equation}
which we diagonalize to determine the single-particle bands given in Eq.~\eqref{band_spec} in the main text.
Note, that we used spinless fermions in (\ref{TBHstand}) reflecting the spin degeneracy of bands in the absence of spin-orbit coupling.
To diagonalize Eq.~\eqref{TBHstand} we use the transformation
\begin{equation}
U^{-1} H_{TB} U = \text{Diag}(E^-_-,E^+_-,E^-_+,E^+_+) \quad \text{with} \quad U=\frac{1}{2\sqrt{|T_{sd1}|}}\begin{pmatrix} i \sqrt{T_{sd1}^*} & -i \sqrt{T_{sd1}^*} & -i \sqrt{T_{sd1}^*} & i \sqrt{T_{sd1}^*} \\ -i \sqrt{T_{sd1}} & i \sqrt{T_{sd1}} & -i \sqrt{T_{sd1}} & i \sqrt{T_{sd1}} \\ -\sqrt{T_{sd1}^*} & -\sqrt{T_{sd1}^*} & \sqrt{T_{sd1}^*} & \sqrt{T_{sd1}^*} \\  \sqrt{T_{sd1}}  &  \sqrt{T_{sd1}}  & \sqrt{T_{sd1}} & \sqrt{T_{sd1}}\end{pmatrix} 
\end{equation}
and the eigenvalues are given in Eq.~\eqref{band_spec}.

With the orbtial-to-band transformation $U$ we transform the fermionic spinor  $\vec{c} = \begin{pmatrix}
c_{xA}^{\dag}
c_{xB}^{\dag}
c_{yA}^{\dag}
c_{yB}^{\dag}
\end{pmatrix}$
written in the Wannier orbital basis to the spinor $\vec{d} = \begin{pmatrix}
d_{1}^{\dag}
d_{2}^{\dag}
d_{3}^{\dag}
d_{4}^{\dag}
\end{pmatrix}$ written in band basis by a standard linear transformation
\begin{equation}
\vec{c} = U \vec{d}.
\end{equation}
Note, that spinor $\vec{c}$ gains momentum dependence from the transformation $U$. With this now we have an explicit relation between the orbital and the band operators, hence we can rewrite the interaction term in the band basis.
As $U$ depends on momentum, this yields the so-called orbital makeup, i.e. additional momentum-dependent factors (which are sometimes called ``coherence factors'') in the interaction from the transformation.

In our model we are interested in fermions located at patches near the VH  points.
So as a next step, we introduce patch band operators, which ``live'' at the positions of the VH singularities.
More specifically, we introduce patch operators $f_i, f_i^{\dagger}$ with $i=1..12$ being the patch index defined as
\begin{equation}
f_i = d_j (\vec{k}_i), \; f_i^{\dagger} = d_j^{\dagger} (\vec{k}_i),
\end{equation}
where $\vec{k}_i$ are the positions of VH singularities in the Brillouin zone and $j$ is the band index. Note, that $j$ is different for different patches, since the VH singularities are made of different bands.
The coupling constants can then be obtained by an explicit calculation in which the coupling functions  are translated to momentum space and multiplied by four coherence factors related to fermions on the involved patches.

\section{Explicit form of coupling functions}
\label{app:coupl}

In this section  we present the explicit form of coupling functions from Eq. (\ref{inter_k}). In the equations below we use the same vector notation as in Eqs. (\ref{eq_vec1}),(\ref{eq_vec2}),(\ref{eq_vec3}).
 $QQ-$term coupling functions, which depend only on transferred momentum, are given by
\begin{equation}
\begin{gathered}
F_{AAAA} = \left(1+ e^{i \vec{L}_1 (\vec{k}-\vec{q})} + e^{i \vec{L}_2 (\vec{k}-\vec{q})} \right) \left(1+ e^{i \vec{L}_1 (\vec{k'}-\vec{q'})} + e^{i \vec{L}_2 (\vec{k'}-\vec{q'})} \right), \\
F_{ABBA} = \left(1+ e^{i \vec{L}_1 (\vec{k}-\vec{q})} + e^{i \vec{L}_2 (\vec{k}-\vec{q})} \right) \left( e^{i \vec{a_3} (\vec{k'}-\vec{q'})} + e^{i (\vec{L}_2 + \vec{a_2}) (\vec{k'}-\vec{q'})} +  e^{i (\vec{L}_1 + \vec{a_1}) (\vec{k'}-\vec{q'})} \right), \\
F_{BAAB} =  \left( e^{i \vec{a}_3 (\vec{k}-\vec{q})} + e^{i (\vec{L}_2 + \vec{a}_2) (\vec{k}-\vec{q})} +  e^{i (\vec{L}_1 + \vec{a}_1) (\vec{k}-\vec{q})} \right) \left(1+ e^{i \vec{L}_1 (\vec{k'}-\vec{q'})} + e^{i \vec{L}_2 (\vec{k'}-\vec{q'})} \right), \\
F_{BBBB} =  \left( e^{i \vec{a}_3 (\vec{k}-\vec{q})} + e^{i (\vec{L}_2 + \vec{a}_2) (\vec{k}-\vec{q})} +  e^{i (\vec{L}_1 + \vec{a}_1) (\vec{k}-\vec{q})} \right)  \left( e^{i \vec{a}_3 (\vec{k'}-\vec{q'})} + e^{i (\vec{L}_2 + \vec{a}_2) (\vec{k'}-\vec{q'})} +  e^{i (\vec{L}_1 + \vec{a}_1) (\vec{k'}-\vec{q'})} \right).
\end{gathered}
\label{FQQ}
\end{equation}

Couplings of $TQQT-$ and $TT-$terms can also be written in an explicit form. For $TQQT$ we get
\begin{multline}
F_{AABA} =  \left(1+ e^{i \vec{L}_1 (\vec{k}-\vec{q})} + e^{i \vec{L}_2 (\vec{k}-\vec{q})} \right) \Big( e^{-i \vec{\delta}_3 \vec{q'}} + e^{i \vec{L}_1 \vec{k'}} e^{-i (\vec{L}_1 + \vec{a}_1) \vec{q'}} + e^{i \vec{L}_2 \vec{k'}} e^{-i (\vec{L}_2 + \vec{a}_2) \vec{q'}} - e^{i \vec{L}_1 \vec{k'}} e^{-i \vec{a}_3 \vec{q'}} - e^{i \vec{L}_2 \vec{k'}} e^{-i (\vec{L}_1 + \vec{a}_1) \vec{q'}} -  e^{-i (\vec{L}_2 + \vec{a}_2) \vec{q'}} \Big), \\
F_{ABAA} = \left(1+ e^{i \vec{L}_1 (\vec{k}-\vec{q})} + e^{i \vec{L}_2 (\vec{k}-\vec{q})} \right) \Big( -e^{-i \vec{a}_3 \vec{k'}} e^{i \vec{L}_1 \vec{q'}} - e^{i (\vec{L}_1 + \vec{a}_1) \vec{k'}} e^{-i \vec{L}_2 \vec{q'}}  - e^{-i (\vec{L}_2 + \vec{a}_2) \vec{k'}}+ e^{i \vec{a}_3 \vec{k'}}  +  e^{i (\vec{L}_1 + \vec{a}_1) \vec{k'}} e^{-i \vec{L}_1 \vec{q'}} +  e^{i (\vec{L}_2 + \vec{a}_2) \vec{k'}} e^{-i \vec{L}_2 \vec{q'}} \Big), \\
F_{AAAB} = F_{AABA}(\vec{k} \leftrightarrow \vec{k'}, \vec{q} \leftrightarrow \vec{q'}), \\
F_{BAAA} = F_{ABAA}(\vec{k} \leftrightarrow \vec{k'}, \vec{q} \leftrightarrow \vec{q'}), \\
F_{BABB} = \left( e^{i \vec{a_3} (\vec{k}-\vec{q})} + e^{i (\vec{L}_2 + \vec{a_2}) (\vec{k}-\vec{q})} +  e^{i (\vec{L}_1 + \vec{a_1})  (\vec{k}-\vec{q})} \right) \Big( e^{-i \vec{a}_3 \vec{q'}} + e^{i \vec{L}_1 \vec{k'}} e^{-i (\vec{L}_1 + \vec{a}_1) \vec{q'}} + e^{i \vec{L}_2 \vec{k'}} e^{-i (\vec{L}_2 + \vec{a}_2) \vec{q'}}-  e^{i \vec{L}_1 \vec{k'}} e^{-i \vec{a}_3 \vec{q'}} - e^{i \vec{L}_2 \vec{k'}} e^{-i (\vec{L}_1 + \vec{a}_1) \vec{q'}} -  e^{-i (\vec{L}_2 + \vec{a}_2) \vec{q'}} \Big), \\
F_{BBAB} =  \left( e^{i \vec{a_3} (\vec{k}-\vec{q})} + e^{i (\vec{L}_2 + \vec{a_2}) (\vec{k}-\vec{q})} +  e^{i (\vec{L}_1 + \vec{a_1}) (\vec{k}-\vec{q})} \right) \Big( -e^{-i \vec{a}_3 \vec{k'}} e^{i \vec{L}_1 \vec{q'}} - e^{i (\vec{L}_1 + \vec{a}_1) \vec{k'}} e^{-i \vec{L}_2 \vec{q'}} -  e^{-i (\vec{L}_2 + \vec{a}_2) \vec{k'}}+  e^{i \vec{a}_3 \vec{k'}}  +  e^{i (\vec{L}_1 + \vec{a}_1) \vec{k'}} e^{-i \vec{L}_1 \vec{q'}} +  e^{i (\vec{L}_2 + \vec{a}_2) \vec{k'}} e^{-i \vec{L}_2 \vec{q'}} \Big), \\
F_{ABBB} = F_{BABB}(\vec{k} \leftrightarrow \vec{k'}, \vec{q} \leftrightarrow \vec{q'}), \\
F_{BBBA} = F_{BBAB}(\vec{k} \leftrightarrow \vec{k'}, \vec{q} \leftrightarrow \vec{q'}), \\
\label{FTQQT}
\end{multline}
while  $TT-$couplings are
\begin{align}
F_{ABAB} &= \Big( e^{-i \vec{a}_3 \vec{q}} + e^{i \vec{L}_1 \vec{k}} e^{-i (\vec{L}_1 + \vec{a}_1) \vec{q}}+ e^{i \vec{L}_2 \vec{k}} e^{-i (\vec{L}_2 + \vec{a}_2) \vec{q'}}-  e^{i \vec{L}_1 \vec{k}} e^{-i \vec{a}_3 \vec{q}}- e^{i \vec{L}_2 \vec{k}} e^{-i (\vec{L}_1 + \vec{a}_1) \vec{q}} -   e^{-i (\vec{L}_2 + \vec{a}_2) \vec{q}} \Big) \notag\\
&\times\Big( -e^{-i \vec{a}_3 \vec{k'}} e^{i \vec{L}_1 \vec{q'}} - e^{i (\vec{L}_1 + \vec{a}_1) \vec{k'}} e^{-i \vec{L}_2 \vec{q'}}  - e^{-i (\vec{L}_2 + \vec{a}_2) \vec{k'}}+  e^{i \vec{a}_3 \vec{k'}}  +  e^{i (\vec{L}_1 + \vec{a}_1) \vec{k'}} e^{-i \vec{L}_1 \vec{q'}} +  e^{i (\vec{L}_2 + \vec{a}_2) \vec{k'}} e^{-i \vec{L}_2 \vec{q'}} \Big), \\
F_{AABB} &= \Big( e^{-i \vec{a}_3 \vec{q}} + e^{i \vec{L}_1 \vec{k}} e^{-i (\vec{L}_1 + \vec{a}_1) \vec{q}}+ e^{i \vec{L}_2 \vec{k}} e^{-i (\vec{L}_2 + \vec{a}_2) \vec{q'}}-  e^{i \vec{L}_1 \vec{k}} e^{-i \vec{a}_3 \vec{q}} - e^{i \vec{L}_2 \vec{k}} e^{-i (\vec{L}_1 + \vec{a}_1) \vec{q}} -   e^{-i (\vec{L}_2 + \vec{a}_2) \vec{q}} \Big) \notag\\
&\times \Big( e^{-i \vec{a}_3 \vec{q'}} + e^{i \vec{L}_1 \vec{k'}} e^{-i (\vec{L}_1 + \vec{a}_1) \vec{q'}} + e^{i \vec{L}_2 \vec{k'}} e^{-i (\vec{L}_2 + \vec{a}_2) \vec{q'}} - e^{i \vec{L}_1 \vec{k'}} e^{-i \vec{a}_3 \vec{q'}} - e^{i \vec{L}_2 \vec{k'}} e^{-i (\vec{L}_1 + \vec{a}_1) \vec{q'}} -  e^{-i (\vec{L}_2 + \vec{a}_2) \vec{q'}} \Big), \\
F_{BABA} &= \Big( -e^{-i \vec{a}_3 \vec{k}} e^{i \vec{L}_1 \vec{q}} - e^{i (\vec{L}_1 + \vec{a}_1) \vec{k}} e^{-i \vec{L}_2 \vec{q}} - e^{-i (\vec{L}_2 + \vec{a}_2) \vec{k}}+  e^{i \vec{a}_3 \vec{k}}  + e^{i (\vec{L}_1 + \vec{a}_1) \vec{k}} e^{-i \vec{L}_1 \vec{q}} +  e^{i (\vec{L}_2 + \vec{a}_2) \vec{k}} e^{-i \vec{L}_2 \vec{q}} \Big) \notag\\
&\times\Big( e^{-i \vec{a}_3 \vec{q'}} + e^{i \vec{L}_1 \vec{k'}} e^{-i (\vec{L}_1 + \vec{a}_1) \vec{q'}} + e^{i \vec{L}_2 \vec{k'}} e^{-i (\vec{L}_2 + \vec{a}_2) \vec{q'}} -  e^{i \vec{L}_1 \vec{k'}} e^{-i \vec{a}_3 \vec{q'}} - e^{i \vec{L}_2 \vec{k'}} e^{-i (\vec{L}_1 + \vec{a}_1) \vec{q'}} - e^{-i (\vec{L}_2 + \vec{a}_2) \vec{q'}} \Big), \\
F_{BBAA} &= \Big( -e^{-i \vec{a}_3 \vec{k}} e^{i \vec{L}_1 \vec{q}} - e^{i (\vec{L}_1 + \vec{a}_1) \vec{k}} e^{-i \vec{L}_2 \vec{q}} - e^{-i (\vec{L}_2 + \vec{a}_2) \vec{k}}+  e^{i \vec{a}_3 \vec{k}} + e^{i (\vec{L}_1 + \vec{a}_1) \vec{k}} e^{-i \vec{L}_1 \vec{q}} +  e^{i (\vec{L}_2 + \vec{a}_2) \vec{k}} e^{-i \vec{L}_2 \vec{q}} \Big) \notag \\
&\times\Big( -e^{-i \vec{a}_3 \vec{k'}} e^{i \vec{L}_1 \vec{q'}} - e^{i (\vec{L}_1 + \vec{a}_1) \vec{k'}} e^{-i \vec{L}_2 \vec{q'}}  - e^{-i (\vec{L}_2 + \vec{a}_2) \vec{k'}}+  e^{i \vec{a}_3 \vec{k'}}  +  e^{i (\vec{L}_1 + \vec{a}_1) \vec{k'}} e^{-i \vec{L}_1 \vec{q'}} +  e^{i (\vec{L}_2 + \vec{a}_2) \vec{k'}} e^{-i \vec{L}_2 \vec{q'}} \Big).
\label{FTT}
\end{align}

\section{Hubbard-Stratonovich derivation and the symmetry properties of the Landau free energy}
\label{app:HS_GL}

In this section  we provide an explicit derivation of the Landau free energy  using the Hubbard-Stratonovich transformation and discuss symmetry properties of the functional.
We first derive the free energy.
The derivation for the six-patch model is analogous to the calculation presented in the Supplementary material for \cite{Nandkishore2012}.
Here, we present the derivation for the twelve-patch model. We begin by writing down the Lagrangian
\begin{equation}
L = \sum_i f_i^{\dag} (\partial_{\tau} -\epsilon_k^i) f_i -
\begin{pmatrix}
f_1^{\dag} f_1^{\dag} & f_2^{\dag} f_2^{\dag} & f_3^{\dag} f_3^{\dag} & f_4^{\dag} f_4^{\dag} & f_5^{\dag} f_5^{\dag} & f_6^{\dag} f_6^{\dag}
\end{pmatrix}
\begin{pmatrix}

v_0 & g_{1-} & g_2 & g_3 & g_2 & g_{1+} \\
g_{1-} & v_0 & g_{1+} &  g_2 & g_3 & g_2  \\
g_2 & g_{1+} & v_0 & g_{1-} & g_2 & g_3 \\
g_3 & g_2 & g_{1-} & v_0 & g_{1+} & g_2 \\
g_2 & g_3 & g_2 & g_{1+} & v_0 & g_{1-} \\
g_{1+} & g_2 & g_3 & g_2 & g_{1-} & v_0
\end{pmatrix}
\begin{pmatrix}
f_1 f_1 \\ f_2 f_2 \\ f_3 f_3 \\ f_4 f_4 \\ f_5 f_5 \\ f_6 f_6
\end{pmatrix},
\label{Lagr}
\end{equation}
where we absorbed the chemical potential $\mu$ into the dispersion $\epsilon_k$, $i$ - is the patch index running from $1$ to $6$ and the other indices were omitted for shortness.
The spin and band structure of each term is $f_{a\tau\sigma}^\dag f_{a\bar\tau\bar\sigma}^\dag f_{a\bar\tau\bar\sigma} f_{a\tau\sigma}$, where the bar labels opposite spin or band, i.e. we consider spin-singlet pairing with zero total momentum.
Note, that due to the $D_{3}$ symmetry of our low-energy model fermionic dispersions are identical for all patches (upon rotations).
We decompose the pairing interaction in Eq.~\eqref{Lagr} into its eigenvalues and eigenvectors. A pairing instability can develop when at least one eigenvalue is positive and the dominant pairing channel is determined by the largest eigenvalue. We find that two eigenvalues can become positive. They belong to either the $A_2$ or $E$ representation of $D_{3}$. As $A_2$ is a one-dimensional irreducible representation, the corresponding eigenvalue is unique, while the eigenvalue of the two-dimensional $E$ representation is two-fold degenerate.
We discard subleading channels in the following, which is justified close to the largest critical temperature because critical temperatures of subleading channels are much smaller.
In contrast, the critical temperatures of $A_2$ and $E$ order are very similar.
We can classify the eigenvectors corresponding to the leading pairing instabilities according to the lattice harmonics.
Interestingly, the eigenvectors appear in higher ``angular'' momentum channels in our case following $g_{E_1}(k_x,k_y)=k_x^4-6k_x^2k_y^2+k_y^4=\cos 4\theta$, $g_{E_2}(k_x,k_y)=4k_xk_y(k_x^2-k_y^2)=\sin 4\theta$ and $g_{A_2}(k_x,k_y)=6k_x^5k_y-20k_x^3k_y^3+6k_xk_y^5=\sin 6\theta$ with $\theta=\arctan k_y/k_x$. That is we find the eigenvectors
\begin{align}
\vec\Delta_{12p}^{A_2}&=\left(g_{A_2}(\vec k_1),\ldots,g_{A_2}(\vec k_6)\right)\notag\\
\vec\Delta_{12p}^{E_1^-}&=\left(g_{E_1}(\vec k_1),\ldots,g_{E_1}(\vec k_6)\right)\notag \\
\vec\Delta_{12p}^{E_2^-}&=\left(g_{E_2}(\vec k_1),\ldots,g_{E_2}(\vec k_6)\right)
\label{evs}
\end{align}
with the six VH points $\vec k_1,\ldots, \vec k_6$. In the following, we will use the orthonormal expressions $\vec\Delta_{12p}^{\Gamma}\rightarrow \vec\Delta_{12p}^{\Gamma}/\lVert \vec\Delta_{12p}^{\Gamma} \rVert$.
Let us also note that we can choose any linear, orthonormal combination of $\vec\Delta_{12p}^{E_1^-}$ and $\vec\Delta_{12p}^{E_2^-}$ in the two-dimensional subspace corresponding to the degenerate eigenvalue, i.e. $\vec\Delta_{12p}^{E_1^-} \to \cos(\alpha) \vec\Delta_{12p}^{E_1^-} - \sin(\alpha) \vec\Delta_{12p}^{E_2^-}, \vec\Delta_{12p}^{E_2^-} \to \sin(\alpha) \vec\Delta_{12p}^{E_1^-} + \cos(\alpha) \vec\Delta_{12p}^{E_2^-}$. This corresponds to rotation within the $E$-subgroup and the resulting eigenvectors with proper normalization can be represented as
\begin{align}
\vec{\Delta}_{12p}^{A_2}&= N^{-1}_{A_2}  \big(\sin (6\theta_1),\ldots ,\sin (6\theta_6)\big)^T \notag\\
\vec{\Delta}_{12p}^{E_1^-}&= N^{-1}_{E_1} \big(\cos (4\theta_1+\alpha),\ldots ,\cos (4\theta_6+\alpha)\big)^T \notag\\
\vec{\Delta}_{12p}^{E_2^-}&= N^{-1}_{E_2} \big(\sin (4\theta_1+\alpha),\ldots ,\sin (4\theta_6+\alpha)\big)^T,
\label{evs2}
\end{align}
with $N^{-1}_{A,E}$ being the normalization coefficients.
 Explicitly, for a particular choice of hopping parameters (see caption for Fig. \ref{fig_bands}), by numerical diagonalization of the linearized gap equation we obtain $\vec\Delta_{12p}^{A_2}=  \frac{1}{\sqrt{6}} (-1,1,-1,1,-1,1), \vec\Delta_{12p}^{E_1^-} = (-0.273,  0.018, 0.577, -0.509, -0.304, 0.491)$ and $\vec\Delta_{12p}^{E_2^-} = (0.509, -0.577, -0.018, 0.273, -0.491, 0.304).$ To an excellent approximation these gap values are fitted by single harmonics given in Eq. \ref{evs2} with $\alpha=3\pi/4$. We use these three vectors further to perform the Hubbard-Stratonovich transformation.

In the bulk text we have provided a solution which shows that at a critical value of the parameter $\alpha_T$, all three channels become degenerate.
Close to this value, the pairing function can be expressed as a linear combination of the eigenvectors
\begin{equation}
\Delta^{SC}_{12p}=\Delta_{A_2}\vec\Delta_{12p}^{A_2} + \Delta_{E_1}\vec{\Delta}_{12p}^{E_1^-} + \Delta_{E_2} \vec{\Delta}_{12p}^{E_2^-}.
\end{equation}
We are interested in the Landau action in the vicinity of exactly this point, which can be constructed as an expansion in $\Delta_{A_2},\Delta_{E_1}$ and $\Delta_{E_2}$. As we said in the main text, we can anticipate the form of the Landau action from symmetry properties; it has to be invariant under $U(1)$ and $C_3$ transformations. At the same time, we can derive it explicitly from the fermionic Lagrangian, which also gives us the bare values of the couplings in the Landau action.

For the following derivation, it is convenient to
introduce three $6 \times 6$ matrices $ d_{A_2} ,  d_{E_1} ,$ and $ d_{E_2} $ in patch space. These matrices are diagonal and given by
\begin{equation}
d_{A_2}= \mathrm{diag} (\vec\Delta_{12p}^{A_2}) \qquad d_{E_1}= \mathrm{diag} (\vec{\Delta}_{12p}^{E_1^-} ) \qquad d_{E_2}=\mathrm{diag} (\vec{\Delta}_{12p}^{E_2^-})
\end{equation}
The order parameters then can be defined via these matrices
\begin{equation}
\begin{gathered}
\Delta_{A_2} = 2 \lambda \av{f^T d_{A_2} f},\\
\Delta_{E1} = 2 \lambda \av{f^T d_{E_1} f},\\
\Delta_{E2} = 2 \lambda \av{f^T d_{E_2} f},
\end{gathered}
\end{equation}
where $f^T=(f_1,\ldots,f_6)$ and $\lambda$  is the triply degenerate eigenvalue. Using these bosonic fields we can now apply the Hubbard-Stratonovich transformation to Eq. (\ref{Lagr})
and obtain the Lagrangian in terms of bosonic and fermionic fields written in the Nambu-Gor'kov notation:
\begin{equation}
L=
\begin{pmatrix}
f^{\dag} & f
\end{pmatrix}
\begin{pmatrix}
G^{-1}_{+} && \Delta_{A_2}  d_{A_2}  + \Delta_{E1}  d_{E_1}  + \Delta_{E2}  d_{E_2}  \\
\Delta_{A_2}^*  d_{A_2}  + \Delta_{E1}^*  d_{E_1}  + \Delta_{E2}^*  d_{E_2}   && G^{-1}_{-}
\end{pmatrix}
\begin{pmatrix}
f \\ f^{\dag}
\end{pmatrix}
+ \frac{|\Delta_{A_2}|^2 + |\Delta_{E1}|^2 + |\Delta_{E2}|^2}{2 \lambda},
\label{Lagr_HS1}
\end{equation}
where $G_{\pm}$ are particle and hole Green's functions with $G_{\pm}=\text{diag}(G_{1,\pm},\ldots,G_{6,\pm}$).
The Green's functions are given by $G_{i,\pm}^{-1} = i \omega \mp \epsilon_k^i,$ where $\omega$  is the Matsubara frequency and $i$ is the patch number index. Now, we can  integrate out the fermions and obtain the Lagrangian written in terms of only bosonic fields
\begin{equation}
L= \Tr \; \ln
\begin{pmatrix}
G^{-1}_+ && \Delta_{A_2}  d_{A_2}  + \Delta_{E1}  d_{E_1}  + \Delta_{E2}  d_{E_2}  \\
\Delta_{A_2}^*  d_{A_2}  + \Delta_{E1}^*  d_{E_1}  + \Delta_{E2}^*  d_{E_2}   && G^{-1}_-
\end{pmatrix}
+ \frac{|\Delta_{A_2}|^2 + |\Delta_{E1}|^2 + |\Delta_{E2}|^2}{2 \lambda},
\label{Lagr_HS2}
\end{equation}
where the trace is going over the patch and Nambu-Gor'kov spinor spaces (the overall matrix dimension in (\ref{Lagr_HS2}) is twelve). The trace over the patch space is an analog of the momentum integration in the continuous notation. We exploit the property that the Green's functions commute with the order parameter matrices and expand Eq. (\ref{Lagr_HS2}) in small $\Delta_{A_2}, \Delta_{E1}, \Delta_{E2}$ up to the fourth order.
As a result, we obtain
\begin{align}
\mathcal{F}&=\mathcal{F}_2 + \mathcal{F}_4\notag\\
\mathcal{F}_2&= A_1 \left(|\Delta_{E1}|^2 + |\Delta_{E2}|^2 \right) + A_2 |\Delta_{A_2}|^2 \notag\\
\mathcal{F}_4 &= K \biggr[ \frac{1}{4} \left( |\Delta_{E1}|^4 + |\Delta_{E2}|^4 + \frac{2}{3}|\Delta_{A_2}|^4 \right) + \frac{1}{3} \left( |\Delta_{E1}|^2 |\Delta_{E2}|^2 + 2 |\Delta_{A_2}|^2 |\Delta_{E2}|^2 + 2 |\Delta_{A_2}|^2 |\Delta_{E1}|^2 \right)  \notag\\
&+ \frac{1}{12} \left( \Delta_{E1}^2 \bar{\Delta}_{E2}^2 + \Delta_{E2}^2 \bar{\Delta}_{E1}^2 + 2 \Delta_{A_2}^2 \bar{\Delta}_{E1}^2 + 2 \Delta_{E1}^2 \bar{\Delta}_{A_2}^2 + 2 \Delta_{A_2}^2 \bar{\Delta}_{E2}^2 + 2 \Delta_{E2}^2 \bar{\Delta}_{A_2}^2 \right)  \notag\\
&+\delta \left[ \left( -2 \Delta_{E1} |\Delta_{E2}|^2 - \bar{\Delta}_{E1} \Delta_{E2}^2  +\Delta_{E1} |\Delta_{E1}|^2   \right) \bar{\Delta}_{A_2} + c.c \right] + \delta_2 \left[ \left( 2 \Delta_{E2} |\Delta_{E1}|^2 + \bar{\Delta}_{E2} \Delta_{E1}^2  -\Delta_{E2} |\Delta_{E2}|^2  \right) \bar{\Delta}_{A_2} + c.c \right],
\label{app_F4HS}
\end{align}
where the upper bar means complex conjugation and $K=\int d^2k/(2\pi)^2 \, G_{i,+}G_{i,-}G_{i,+}G_{i,-}$ is the fermionic box diagram with momentum integration constrained to the vicinity of the patches. Note that because of rotation and time-reversal symmetry the constant $K$ is the same for all patches. Coefficients in Eq. (\ref{app_F4HS}) correspond to the ones mentioned in the main text in the following way: $\beta_1 = 1/6, \beta_2 = 1/12, \beta_3=1/6, \gamma_1 = 2/3, \gamma_2 = 1/6.$ To obtain Eq. (\ref{app_F4HS}) we used the identities $\mathrm{Tr} ( d_{E_1} ^4) = \mathrm{Tr} ( d_{E_2} ^4)=1/4 ,\mathrm{Tr} ( d_{E_1} ^2  d_{E_2} ^2) =\mathrm{Tr} ( d_{E_1}   d_{E_2}   d_{E_1}   d_{E_2} ) =1/12, \mathrm{Tr} ( d_{A_2} ^2  d_{E_1} ^2) =\mathrm{Tr} ( d_{A_2} ^2   d_{E_2} ^2 ) = \mathrm{Tr} ( d_{A_2}   d_{E_1}   d_{A_2}   d_{E_1} ) = \mathrm{Tr} ( d_{A_2}   d_{E_2}   d_{A_2}   d_{E_2} ) =\mathrm{Tr} ( d_{A_2} ^4) =1/6,$ and the fact that most traces over the patch space of the cube of one matrix times the other matrix vanishes. However, there are exceptions for special $C_3$- and $U(1)$-symmetric combinations, which lead to the last two lines in Eq.~\eqref{app_F4HS}. We obtain for their prefactor  $\delta=-\delta_2\approx -0.128$ for the choice of the basis for the two-dimensional representation $E$ as in Eq.~\eqref{evs2} with $\alpha=3\pi/4$.
Note that in contrast to the other coupling constants, $\delta$ and $\delta_2$ depend on a basis change of the two-dimensional representation, because the polynomials cubic in $\Delta_{E_1}, \Delta_{E2}$ also transform non-trivially under a basis rotation. Of course, in total, the action is invariant. Explicitly, if we rotate our basis via
\begin{align}
\begin{pmatrix} \vec{\Delta}_{E1}' \\ \vec{\Delta}_{E2}'\end{pmatrix} = R_\alpha \begin{pmatrix} \vec{\Delta}_{E1} \\ \vec{\Delta}_{E2}\end{pmatrix} \qquad
R_\alpha=\begin{pmatrix} \cos\alpha & -\sin\alpha \\ \sin\alpha & \cos\alpha \end{pmatrix}
\end{align}
the polynomials $p_1= \Delta_{E1} |\Delta_{E1}|^2- 2 \Delta_{E1} |\Delta_{E2}|^2-  \bar{\Delta}_{E1} \Delta_{E2}^2, p_2=-\Delta_{E2} |\Delta_{E2}|^2+ 2 \Delta_{E2} |\Delta_{E1}|^2+  \bar{\Delta}_{E2} \Delta_{E1}^2$ and $\delta,\delta_2$ transform according to
\begin{align}
\begin{pmatrix} \delta' &  \delta_2'\end{pmatrix} =\begin{pmatrix} \delta & \delta_2\end{pmatrix}R_{3\alpha}  \qquad \begin{pmatrix} p_1' \\ p _2'\end{pmatrix} = R_{3\alpha}^T \begin{pmatrix} p_1 \\ p_2\end{pmatrix}
\end{align}
so that their product in $\mathcal{F}$ is invariant. This allows us to choose our basis so that $\delta=-\delta_2$ to simplify our discussion of the Landau action.

\end{document}